\documentclass[11pt]{article}

\usepackage{epsfig,subfigure}
\usepackage{amssymb}
\usepackage[fleqn]{amsmath}
\usepackage{color}
\usepackage{arydshln}

\def\QED{\mbox{\rule[0pt]{1.5ex}{1.5ex}}}
\def\proof{\noindent\hspace{2em}{\it Proof: }}
\def\new{\mbox{\texttt{new}}}

\usepackage{graphicx}
\usepackage{color}
\usepackage[dvipsnames]{xcolor}

\definecolor{armygreen}{rgb}{0.29, 0.33, 0.13}

\usepackage{amssymb}
\usepackage{amsmath,amsfonts,amssymb}
\usepackage{verbatim}
\usepackage{stfloats}
\usepackage[bookmarks=false]{}
\usepackage{algorithm}
\usepackage{algcompatible}

\newtheorem{theorem}{Theorem}

\newtheorem{lemma}{Lemma}

\newcommand\m[1]{%
\mbox{\small #1}%
}

\newcommand\blfootnote[1]{%
  \begingroup
  \renewcommand\thefootnote{}\footnote{#1}%
  \addtocounter{footnote}{-1}%
  \endgroup
}

\begin{document}
\date{}
\title{
The Capacity of Private Information Retrieval
}
\author{ \normalsize Hua Sun and Syed A. Jafar \\
}

\maketitle

\blfootnote{Hua Sun (email: huas2@uci.edu) and Syed A. Jafar (email: syed@uci.edu) are with the Center of Pervasive Communications and Computing (CPCC) in the Department of Electrical Engineering and Computer Science (EECS) at the University of California Irvine. The work is supported by grants from  ONR, NSF and ARL. The results of this work were submitted in part for presentation at IEEE ISIT 2016 and IEEE GLOBECOM 2016. }

\begin{abstract}
In the private information retrieval (PIR) problem a user wishes to retrieve, as efficiently as possible, one out of $K$ messages from $N$ non-communicating databases (each holds all $K$ messages) while revealing nothing about the identity of the desired message index to any individual database. The information theoretic capacity of PIR is the maximum number of bits of desired information that can be privately retrieved per bit of downloaded information. For $K$ messages and $N$ databases, we show that the PIR capacity is $\left(1+1/N+1/N^2+\cdots+1/N^{K-1}\right)^{-1}$. A remarkable feature of the  capacity achieving scheme is that if we eliminate any subset of messages (by setting the message symbols to zero), the resulting scheme also achieves the PIR capacity for the remaining subset of messages.
\end{abstract}

\allowdisplaybreaks
\section{Introduction}
Marked by paradigm-shifting developments such as big data, cloud computing, and internet of things, the modern information age presents researchers with an unconventional set of challenges. The rapidly evolving research landscape continues to blur traditional boundaries between computer science, communication and information theory, coding and signal processing.  For example, the index coding problem which was introduced by computer scientists in 1998 \cite{Birk_Kol, Birk_Kol_Trans}, is now a very active  research topic in information theory because of its fundamental connections to a broad range of questions that includes topological interference management \cite{Jafar_TIM}, network coding \cite{Rouayheb_Sprintson_Georghiades}, distributed storage capacity \cite{Mazumdar}, hat guessing \cite{Riis_Hat}, and non-Shannon information inequalities \cite{Sun_Jafar_nonshannon}. Evidently, the crossover of problems across fields creates exciting opportunities for fundamental progress through a consolidation of complementary perspectives. The pursuit of such crossovers brings us to the private information retrieval (PIR) problem \cite{Yekhanin, William,Sun_Jafar_BIAPIR}. 


Introduced in 1995 by Chor, Kushilevitz, Goldreich and Sudan \cite{PIRfirst, PIRfirstjournal}, the private information retrieval (PIR) problem seeks the most efficient way for a user to retrieve a desired message from a set of distributed databases, each of which stores all the messages, without revealing any information about which message  is being retrieved to any individual database.  The user can hide his interests trivially by requesting all the information, but that could be very inefficient (expensive). The goal of the PIR problem is to find the most efficient solution. 

Besides its direct applications, PIR is of broad interest because it shares intimate connections to many other prominent problems. PIR attracted our attention initially  in \cite{Sun_Jafar_BIAPIR} because of its curious similarities to  Blind Interference Alignment \cite{Jafar_corr}. PIR protocols are the essential ingredients of oblivious transfer \cite{SymPIR}, instance hiding \cite{Hide, Hide_one, Hide_multiple},  multiparty computation \cite{Local_random}, secret sharing schemes \cite{Shamir, Beimel_Ishai_Kushilevitz_Orlov} and locally decodable codes \cite{YekhaninPhd}. Through the connection between locally decodable and locally recoverable codes \cite{Gopalan_Huang_Simitci_Yekhanin}, PIR also connects to  distributed data storage repair \cite{Dimakis_survey}, index coding \cite{Birk_Kol_Trans} and the entire umbrella of  network coding \cite{Ahlswede_Cai_etal} in general.   As such PIR holds tremendous promise as a point of convergence of complementary perspectives. The characterization of the information theoretic capacity of PIR that we undertake in this work, is a step in this direction.

The PIR problem is described as follows.  We have $N$ non-communicating databases, 
each stores the full set of $K$ independent messages $W_1, \cdots, W_K$. A user wants one of the messages, say $W_\theta, \theta \in \{1,2, \cdots, K\}$, but requires each database to learn absolutely nothing (in the information theoretic sense)\footnote{There is another line of research, where privacy needs to be satisfied only for computationally bounded databases \cite{William, Yekhanin, CPIR}.}  about the retrieved message index, $\theta$. To do so, the user generates $N$ queries $Q_1, \cdots, Q_N$ and sends $Q_n, n \in \{1,2,\cdots, N\}$ to the $n$-th database. After receiving query $Q_n$, the $n$-th database returns an answering string $A_n$ to the user. The user must be able to obtain the desired message $W_\theta$ from all the answers $A_1, \cdots, A_N$. To be private, each query $Q_n$ and each answer $A_n$ must be independent of the desired message index, $\theta$. 

For example, suppose we have $N = 2$ databases and $K$ messages. To retrieve $W_\theta$ privately, the user first generates a random length-$K$ vector $[h_1, h_2, \cdots, h_K]$, where each element is independent and identically distributed uniformly over $\mathbb{F}_2$, i.e., equally likely to be $0$ or $1$. Then the user sends $Q_1 = [h_1, h_2, \cdots, h_\theta,\cdots, h_K]$ to the first database and $Q_2 = [h_1, h_2, \cdots,h_{\theta-1}, (h_{\theta}+1),h_{\theta+1},\cdots, h_K]$ to the second database. Each database uses the query vector as the combining coefficients and produces the corresponding linear combination of message bits as the answer to the query.
\begin{eqnarray}
A_1&=&\sum_{k=1}^Kh_kW_k\label{eq:PIR1}\\
A_2&=&\sum_{k=1}^Kh_kW_k+W_\theta\label{eq:PIR2}
\end{eqnarray}
The user  obtains $W_\theta$ by subtracting $A_1$ from $A_2$. Privacy is guaranteed because  each query is independent of the desired message index $\theta$. This is because regardless of the desired message index $\theta$, each of the query vectors $Q_1, Q_2$ is individually comprised of elements that are i.i.d. uniform over $\mathbb{F}_2$. Thus, each database   learns nothing about which message is requested.

The PIR problem was initially studied in the  setting where each message is one bit long \cite{PIRfirst, PIRfirstjournal, Ambainis, Beimel_Ishai_Kushilevitz, Beimel_Ishai_Kushilevitz_Raymond, YekhaninPhd, 2PIR}, where the cost of a PIR scheme is  measured by the total amount of communication between the user and the databases, i.e., the sum of lengths of each query string (upload) and each answering string (download).
However, for the traditional Shannon theoretic formulation, where message size is allowed to be arbitrarily large, the upload cost is negligible compared to the download cost \cite{Chan_Ho_Yamamoto}\footnote{The justification argument (traces back to Proposition 4.1.1 of \cite{PIRfirstjournal}) is that the upload cost does not scale with the message size. This is because we can reuse the original query functions for each part of the message.}.
In this work we adopt the Shannon theoretic formulation, so that we  focus on the download cost, measured relative to the message size. For the example presented above, each message is $1$ bit and we download a total of $2$ bits (one from each database), so that the download cost is 2 bits per message bit. The reciprocal of  download cost is the  rate, i.e., the number of bits of desired information that is privately retrieved per downloaded information bit.  The maximum rate possible for the PIR problem is its information theoretic capacity $C$. For the  example presented earlier, the private information retrieval rate is $\frac{1}{2}$, meaning that 1 bit of desired information is retrieved from every 2 downloaded bits. In general, for arbitrary $N$ and $K$, the best previously known achievable rate for PIR, reported in \cite{Shah_Rashmi_Kannan}, is $1-\frac{1}{N}$. Since $1$ is a trivial upper bound on capacity, we know that $1\geq C\geq 1-\frac{1}{N}$. The bounds present a reasonable approximation of capacity for large number of databases. However, in this work, we seek the \emph{exact} information theoretic capacity $C$ of the PIR problem, for \emph{arbitrary} number of messages $K$ and \emph{arbitrary} number of databases $N$. 

The paper is organized as follows.  Section \ref{sec:model} presents the problem statement. The exact capacity of PIR is characterized in Section \ref{sec:main}. Section \ref{sec:ach} presents a novel PIR scheme, and Section \ref{sec:con} provides the information theoretic converse (i.e., a tight upper bound) to establish its optimality. Section \ref{sec:disc} contains a discussion of the results and we conclude in Section \ref{sec:conc}.

{\it Notation: For a positive integer $Z$, we use the notation $[Z]=\{1,2,\cdots, Z\}$. The notation $X \sim Y$ is used to indicate that $X$ and $Y$ are identically distributed. Define the notation $A_{n_1:n_2}, n_1, n_2 \in \mathbb{Z}$,  as the set $\{A_{n_1}, A_{n_1+1}, \cdots, A_{n_2}\}$ if $n_1 \leq n_2$, and as the null set otherwise.}

\section{Problem Statement}\label{sec:model}
Consider $K$ independent messages $W_1, \cdots, W_K$ of size $L$ bits each. 
\begin{eqnarray}
&& H(W_1, \cdots, W_K) = H(W_1) + \cdots + H(W_K), \label{h1}\\
&& H(W_1) = \cdots = H(W_K) = L. \label{h2}
\end{eqnarray}
There are $N$ databases and each database stores all the messages $W_1, \cdots, W_K$. In PIR a user privately generates $\theta\in[K]$ and wishes to retrieve $W_\theta$ while keeping $\theta$ a secret from each database. Depending on $\theta$, there are $K$ strategies that the user could employ to privately retrieve his desired message.  For example, if $\theta=k$, then in order to retrieve $W_k$, the user employs $N$ queries $Q_1^{[k]}, \cdots, Q_N^{[k]}$. Since the queries are determined by the user with no knowledge of the realizations of the messages, the queries must be independent of the messages, 
\begin{eqnarray}
\forall k\in[K], &&I(W_1, \cdots, W_K; Q_1^{[k]}, \cdots, Q_N^{[k]}) = 0 \label{qwind}.
\end{eqnarray}
The user  sends query $Q_n^{[k]}$ to the $n$-th database. Upon receiving $Q_n^{[k]}$, the $n$-th database generates an answering string $A_n^{[k]}$, which is a  function of $Q_n^{[k]}$ and the data stored (i.e., all messages $W_1, \cdots, W_K$).
\begin{eqnarray}
\forall k\in[K], \forall n\in[N], &&H(A_n^{[k]} | Q_n^{[k]}, W_1, \cdots, W_K) = 0. \label{ansdet}
\end{eqnarray}
Each database returns to the user its answer $A_n^{[k]}$. From all the information that is now available to the user, he must be able to  decode the desired message $W_k$, with probability of error $P_e$. The probability of error must approach zero as the size of each message $L$ approaches infinity\footnote{If $P_e$ is required to be exactly zero, then the $o(L)$ terms can be replaced with $0$.}. From Fano's inequality, we have 
\begin{eqnarray}
\mbox{[Correctness]} ~\frac{1}{L}H(W_k | A_1^{[k]}, \cdots, A_N^{[k]}, {\color{black} Q_1^{[k]}, \cdots, Q_N^{[k]}}) = o(L) \label{corr}
\end{eqnarray}
where $o(L)$ represents any term whose value approaches zero as $L$ approaches infinity.

To protect the user's privacy, the $K$ strategies must be indistinguishable (identically distributed) from the perspective of each database, i.e., the following privacy constraint must be satisfied\footnote{The privacy constraint is equivalently expressed as $I(\theta;Q_n^{[\theta]},A_n^{[\theta]},W_1,W_2,\cdots,W_K)=0$.} $\forall n\in[N],\forall k\in[K]$:
\begin{eqnarray}
\mbox{[Privacy]} ~~~ (Q_n^{[1]}, A_n^{[1]}, W_1, \cdots, W_K) \sim (Q_n^{[k]}, A_n^{[k]}, W_1, \cdots, W_K) \label{privacy}
\end{eqnarray}

The PIR \emph{rate} characterizes how many bits of desired information are retrieved per downloaded bit, and is defined as follows. 
\begin{eqnarray}
R \triangleq \frac{L}{D} \label{eta_def} 
\end{eqnarray}
where $D$ is the expected value (over random queries) of the total number of bits downloaded by the user from all the databases. Note that because of the privacy constraint (\ref{privacy}), the expected  number of downloaded bits for each message must be the same. 

A rate $R$ is said to be $\epsilon$-error achievable if there exists a sequence of PIR schemes, each of rate greater than or equal to $R$, for which $P_e\rightarrow 0$ as $L\rightarrow\infty.$\footnote{Equivalently, for any $\epsilon>0$, there exists a finite $L_\epsilon$ such that $P_e<\epsilon$ for all $L>L_\epsilon$.} The supremum of $\epsilon$-error achievable rates is called the $\epsilon$-error capacity $C_\epsilon$. A stronger (more constrained)  notion of capacity is the zero-error capacity $C_o$, which is the supremum of zero-error achievable rates. A rate $R$ is said to be zero-error achievable if there exists a PIR scheme of rate greater than or equal to $R$ for which $P_e=0$. From the definitions, it is evident that $C_o \leq C_\epsilon$. While in noise-less settings, the two are often the same, in general the inequality can be strict. Our goal is to characterize both the zero-error capacity, $C_o$, and the $\epsilon$-error capacity, $C_\epsilon$, of PIR.

\section{Main Result: Capacity of Private Information Retrieval}\label{sec:main}
Theorem \ref{thm:download} states the main result.
\begin{theorem}\label{thm:download}
For the private information retrieval problem with $K$ messages and $N$ databases, the capacity is
\begin{eqnarray}
C_o = C_\epsilon = \left(1 +1/N + 1/{N^2} + \cdots +1/{N^{K-1}}\right)^{-1}.
\end{eqnarray}
\end{theorem}
The following observations are in order.
\begin{enumerate}
\item For $N>1$ databases, the capacity expression can  be equivalently expressed as $(1-\frac{1}{N})/(1-\left(\frac{1}{N}\right)^K)$.
\item The capacity is  strictly higher than the previously best known achievable rate of $1-1/N$. 
\item The capacity is a strictly decreasing  function of the number of messages, $K$, and when the number of messages approaches infinity, the capacity approaches $1-1/N$. 
\item The  capacity is strictly increasing in the number of databases, $N$. As the number of databases approaches infinity, the capacity approaches 1.  
\item Since the download cost is the reciprocal of the rate,  Theorem \ref{thm:download}  equivalently  characterizes the optimal download cost per message bit as $\left(1 +1/N + 1/{N^2} + \cdots +1/{N^{K-1}}\right)$ bits.
\item The achievability proof for Theorem \ref{thm:download} to be presented in the next section, shows that message size approaching infinity is not necessary to approach capacity. In fact, it suffices to have messages of size equal to any positive integer multiple of $N^K$ bits (or $N^K$ symbols in any finite field) each to achieve a rate exactly equal to capacity, and with zero-error. 

\item The upper bound proof will show that no PIR scheme can achieve a rate higher than capacity with $P_e\rightarrow 0$ as message size $L\rightarrow \infty$. Unbounded message size is essential to the information theoretic formulation of capacity. However, from a practical standpoint, it is natural to ask what this means if the message size is limited. Finding the optimal rate for limited message size remains an open problem in general. However, we note that regardless of message size, $C_o$ (and therefore also $C_\epsilon$) is always an upper bound on zero-error rate.  For arbitrary message size $L$,  a naive extension of our  PIR scheme can  be obtained as follows. Pad zeros to each message, rounding up the message size to an integer multiple of $N^K$. Then over each block of $N^K$ symbols per message, directly use the capacity achieving PIR scheme. This  achieves the rate $C_o\frac{L}{N^K}/\left\lceil \frac{L}{N^K}\right\rceil $, which matches capacity exactly if $L$ is a positive integer multiple of $N^K$, and otherwise, approaches capacity for large $L$. It is also clearly sub-optimal in general, especially for smaller message sizes where much better schemes are already known. Additional discussion on message size reduction for a capacity achieving PIR scheme is presented in Section \ref{sec:disc}.
\end{enumerate}

\section{{\color{black}Theorem \ref{thm:download}: Achievability}}\label{sec:ach}
We  present a zero-error PIR scheme for $L=N^K$ bits per message in this section, whose rate is equal to capacity. Note that a zero-error scheme with finite message length can always be repeatedly applied to create a sequence of schemes with message-lengths approaching infinity for which the probability of error approaches (is) zero. Thus, the same scheme will suffice as the proof of achievability for both zero-error and $\epsilon$-error capacity.

Let us illustrate the intuition behind the achievable scheme with a few simple examples. Then, based on the examples, we will present an algorithmic description of the achievable scheme for arbitrary number of messages, $K$ and arbitrary number of databases, $N$. We will then revisit the examples in light of the algorithmic formulation. Finally, we will prove that the scheme is both correct and private, and that its rate is equal to the capacity.

\subsection{Two Examples to Illustrate the Key Ideas}
The  capacity achieving PIR scheme has a myopic or greedy character, in that it starts with a narrow focus on the retrieval of the desired message bits from the first database, but grows into a full fledged scheme based on iterative application of three principles: 
\begin{enumerate}
\item[(1)] {\it Enforcing Symmetry Across Databases}
\item[(2)] {\it Enforcing Message Symmetry within the Query to Each Database}
\item[(3)] {\it Exploiting  Side Information of Undesired Messages  to Retrieve New Desired Information}
\end{enumerate}

\subsubsection{Example 1: $N=2, K=2$}
Consider the simplest PIR setting, with $N=2$ databases, and $K=2$ messages with $L = N^K = 4$ bits per message. 
Let $[a_1, a_2, a_3, a_4]$ represent a random permutation of $L = 4$ bits from $W_1$. Similarly, let $[b_1, b_2, b_3, b_4]$ represent an independent random permutation of $L=4$ bits from $W_2$.  These permutations are generated privately and uniformly by the user.

Suppose the desired message is $W_1$, i.e., $\theta = 1$. We start with a query that requests the first bit $a_1$ from the first database (DB1). Applying database symmetry, we  simultaneously request $a_2$ from the second database (DB2). Next, we enforce message symmetry, by including queries for $b_1$ and $b_2$ as the counterparts for $a_1$ and $a_2$. Now we have side information of $b_2$ from DB2 to be exploited in an additional query to DB1, which requests a new desired information bit $a_3$ mixed with $b_2$. Finally, applying database symmetry we have the corresponding query $a_4+b_1$ for DB2. At this point the queries satisfy symmetry across databases, message symmetry within the query to each database, and all undesired side information is exploited, so the construction is complete. The process is explained below, where the number above an arrow indicates which of the three principles highlighted above is  used in each step.
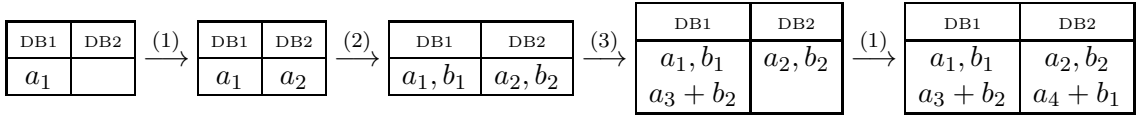
\begin{eqnarray*}
\begin{array}{|c|c|c|c|c|}\hline
\mbox{\tiny DB1}&\mbox{\tiny DB2}\\\hline
a_1&~~\\\hline
\end{array}\stackrel{(1)}{\longrightarrow}\begin{array}{|c|c|c|c|c|}\hline
\mbox{\tiny DB1}&\mbox{\tiny DB2}\\\hline
a_1&a_2\\\hline
\end{array}\stackrel{(2)}{\longrightarrow}\begin{array}{|c|c|c|c|c|}\hline
\mbox{\tiny DB1}&\mbox{\tiny DB2}\\\hline
a_1,b_1&a_2,b_2\\\hline
\end{array}\stackrel{(3)}{\longrightarrow}\begin{array}{|c|c|c|c|c|}\hline
\mbox{\tiny DB1}&\mbox{\tiny DB2}\\\hline
a_1,b_1&a_2,b_2\\
a_3+b_2&\\\hline
\end{array}\stackrel{(1)}{\longrightarrow}\begin{array}{|c|c|c|c|c|}\hline
\mbox{\tiny DB1}&\mbox{\tiny DB2}\\\hline
a_1,b_1&a_2,b_2\\
a_3+b_2&a_4+b_1\\\hline
\end{array}
\end{eqnarray*}

Similarly, the queries for $\theta = 2$ are constructed as follows.
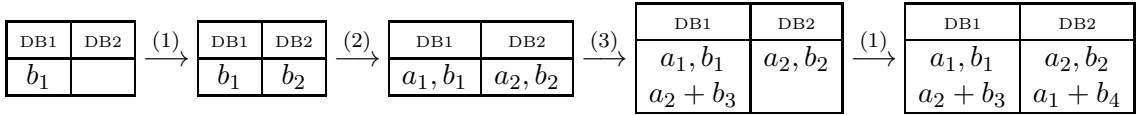
\begin{eqnarray*}
\begin{array}{|c|c|c|c|c|}\hline
\mbox{\tiny DB1}&\mbox{\tiny DB2}\\\hline
b_1&~~\\\hline
\end{array}\stackrel{(1)}{\longrightarrow}\begin{array}{|c|c|c|c|c|}\hline
\mbox{\tiny DB1}&\mbox{\tiny DB2}\\\hline
b_1&b_2\\\hline
\end{array}\stackrel{(2)}{\longrightarrow}\begin{array}{|c|c|c|c|c|}\hline
\mbox{\tiny DB1}&\mbox{\tiny DB2}\\\hline
a_1,b_1&a_2,b_2\\\hline
\end{array}\stackrel{(3)}{\longrightarrow}\begin{array}{|c|c|c|c|c|}\hline
\mbox{\tiny DB1}&\mbox{\tiny DB2}\\\hline
a_1,b_1&a_2,b_2\\
a_2+b_3&\\\hline
\end{array}\stackrel{(1)}{\longrightarrow}\begin{array}{|c|c|c|c|c|}\hline
\mbox{\tiny DB1}&\mbox{\tiny DB2}\\\hline
a_1,b_1&a_2,b_2\\
a_2+b_3&a_1+b_4\\\hline
\end{array}
\end{eqnarray*}


Privacy is ensured by noting that $[a_1,a_2,a_3,a_4]$ is a random permutation of $W_1$ and $[b_1,b_2,b_3,b_4]$ is an independent random permutation of $W_2$. These permutations are only known to the user and not to the databases. Therefore, regardless of the desired message, each database is asked for one randomly chosen bit of each message and a sum of a different pair of randomly chosen bits from each message. Since the permutations are uniform, all possible realizations are equally likely, and privacy is guaranteed.

To verify correctness, note that every desired bit is either downloaded directly or added with known side information which can be subtracted to retrieve the desired bit value. Thus, the desired message bits are successfully recoverable from the downloaded information.

Now, consider the rate of this scheme. The total number of downloaded bits is $6$ and the number of desired bits is $4$. Thus, the rate of this scheme is $4/6=2/3$ which matches the capacity for this case. 

Finally, let us represent the structure of the queries (to any database) in the following  matrix.\begin{eqnarray*}
\begin{array}{r|c|}\cline{2-2}
&\underline{a}\\
&\underline{b}\\
&\underline{a}+\underline{b}\\[0.05cm]
\cline{2-2}
\end{array}
\end{eqnarray*}
 $\underline{a}$ ($\underline{b}$) represents a place-holder for a distinct element of $a_i$ ($b_j$). 
The key to the structure  is that it is made up of sums (a single variable is also named a (trivial) sum) of message bits,  no message bit appears more than once, and all possible assignments of message bits to these place-holders are equally likely. The structure matrix will be useful for the algorithmic description later.

\subsubsection{Example 2: $N=3, K=3$}
The second example is when $N = 3$, $K = 3$. In this case, all messages have $L = N^K = 27$ bits.  
The construction of the optimal PIR scheme for $N=3, K=3$ is illustrated below, where $[a_1, \cdots, a_{27}], [b_1, \cdots, b_{27}], [c_1, \cdots, c_{27}]$ are three i.i.d. uniform permutations of bits from $W_1, W_2, W_3$, respectively. 
The construction of the queries from each database when $\theta = 1$ may be visualized as follows. 
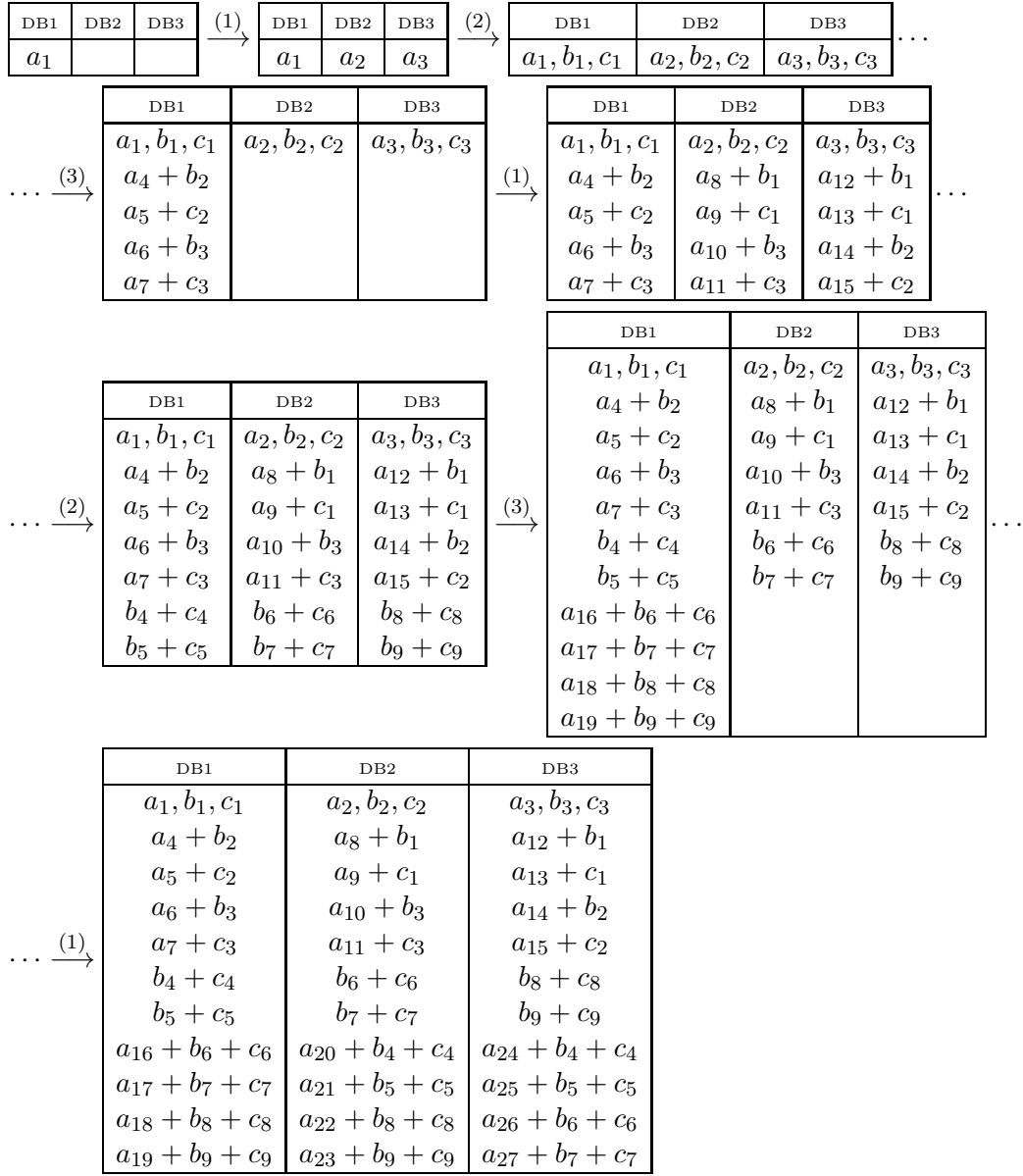
\begin{eqnarray*}
&&\begin{array}{|c|c|c|c|c|c|}\hline
\mbox{\tiny DB1}&\mbox{\tiny DB2}&\mbox{\tiny DB3}\\\hline
a_1&&~~\\\hline
\end{array}\stackrel{(1)}{\longrightarrow}\begin{array}{|c|c|c|c|c|}\hline
\mbox{\tiny DB1}&\mbox{\tiny DB2}&\mbox{\tiny DB3}\\\hline
a_1&a_2&a_3\\\hline
\end{array}\stackrel{(2)}{\longrightarrow}\begin{array}{|c|c|c|c|c|}\hline
\mbox{\tiny DB1}&\mbox{\tiny DB2}&\mbox{\tiny DB3}\\\hline
a_1,b_1,c_1&a_2,b_2,c_2&a_3, b_3, c_3\\\hline
\end{array}\cdots\\
&&\cdots\stackrel{(3)}{\longrightarrow}\begin{array}{|c|c|c|c|c|}\hline
\mbox{\tiny DB1}&\mbox{\tiny DB2}&\mbox{\tiny DB3}\\\hline
a_1,b_1,c_1&a_2,b_2,c_2&a_3, b_3, c_3\\
a_4+b_2&&\\
a_5+c_2&&\\
a_6+b_3&&\\
a_7+c_3&&\\\hline
\end{array}
\stackrel{(1)}{\longrightarrow}\begin{array}{|c|c|c|c|c|}\hline
\mbox{\tiny DB1}&\mbox{\tiny DB2}&\mbox{\tiny DB3}\\\hline
a_1,b_1,c_1&a_2,b_2,c_2&a_3, b_3, c_3\\
a_4+b_2&a_8+b_1&a_{12}+b_1\\
a_5+c_2&a_9+c_1&a_{13}+c_1\\
a_6+b_3&a_{10}+b_3&a_{14}+b_2\\
a_7+c_3&a_{11}+c_3&a_{15}+c_2\\\hline
\end{array}\cdots\\
&&\cdots\stackrel{(2)}{\longrightarrow}\begin{array}{|c|c|c|c|c|}\hline
\mbox{\tiny DB1}&\mbox{\tiny DB2}&\mbox{\tiny DB3}\\\hline
a_1,b_1,c_1&a_2,b_2,c_2&a_3, b_3, c_3\\
a_4+b_2&a_8+b_1&a_{12}+b_1\\
a_5+c_2&a_9+c_1&a_{13}+c_1\\
a_6+b_3&a_{10}+b_3&a_{14}+b_2\\
a_7+c_3&a_{11}+c_3&a_{15}+c_2\\
b_4+c_4&b_{6}+c_6&b_{8}+c_8\\
b_5+c_5&b_{7}+c_7&b_{9}+c_9\\
\hline
\end{array}\stackrel{(3)}{\longrightarrow}\begin{array}{|c|c|c|c|c|}\hline
\mbox{\tiny DB1}&\mbox{\tiny DB2}&\mbox{\tiny DB3}\\\hline
a_1,b_1,c_1&a_2,b_2,c_2&a_3, b_3, c_3\\
a_4+b_2&a_8+b_1&a_{12}+b_1\\
a_5+c_2&a_9+c_1&a_{13}+c_1\\
a_6+b_3&a_{10}+b_3&a_{14}+b_2\\
a_7+c_3&a_{11}+c_3&a_{15}+c_2\\
b_4+c_4&b_{6}+c_6&b_{8}+c_8\\
b_5+c_5&b_{7}+c_7&b_{9}+c_9\\
a_{16} + b_{6}+c_6&&\\
a_{17} + b_{7}+c_7&&\\
a_{18} + b_{8}+c_8&&\\
a_{19} + b_{9}+c_9&&\\
\hline
\end{array}\cdots\\
&&\cdots\stackrel{(1)}{\longrightarrow}\begin{array}{|c|c|c|c|c|}\hline
\mbox{\tiny DB1}&\mbox{\tiny DB2}&\mbox{\tiny DB3}\\\hline
a_1,b_1,c_1&a_2,b_2,c_2&a_3, b_3, c_3\\
a_4+b_2&a_8+b_1&a_{12}+b_1\\
a_5+c_2&a_9+c_1&a_{13}+c_1\\
a_6+b_3&a_{10}+b_3&a_{14}+b_2\\
a_7+c_3&a_{11}+c_3&a_{15}+c_2\\
b_4+c_4&b_{6}+c_6&b_{8}+c_8\\
b_5+c_5&b_{7}+c_7&b_{9}+c_9\\
a_{16} + b_{6}+c_6&a_{20} + b_{4}+c_4&a_{24} + b_{4}+c_4\\
a_{17} + b_{7}+c_7&a_{21} + b_{5}+c_5&a_{25} + b_{5}+c_5\\
a_{18} + b_{8}+c_8&a_{22} + b_{8}+c_8&a_{26} + b_{6}+c_6\\
a_{19} + b_{9}+c_9&a_{23} + b_{9}+c_9&a_{27} + b_{7}+c_7\\
\hline
\end{array}
\end{eqnarray*}

Similarly, the queries when $\theta = 2, 3$ are as follows.
\begin{eqnarray*}
\begin{array}{ccc}
\theta = 2 &~~& \theta = 3\\
\begin{array}{|c|c|c|c|c|}\hline
\mbox{\tiny DB1}&\mbox{\tiny DB2}&\mbox{\tiny DB3}\\\hline
a_1,b_1,c_1&a_2,b_2,c_2&a_3, b_3, c_3\\
a_2+b_4&a_1+b_8&a_1+b_{12}\\
b_5+c_2&b_9+c_1&b_{13}+c_1\\
a_3+b_6&a_3+b_{10}&a_2+b_{14}\\
b_7+c_3&c_3+b_{11}&b_{15}+c_2\\
a_4+c_4&a_{6}+c_6&a_{8}+c_8\\
a_5+c_5&a_{7}+c_7&a_{9}+c_9\\
a_{6} + b_{16}+c_6&a_{4} + b_{20}+c_4&a_{4} + b_{24}+c_4\\
a_{7} + b_{17}+c_7&a_{5} + b_{21}+c_5&a_{5} + b_{25}+c_5\\
a_{8} + b_{18}+c_8&a_{8} + b_{22}+c_8&a_{6} + b_{26}+c_6\\
a_{9} + b_{19}+c_9&a_{9} + b_{23}+c_9&a_{7} + b_{27}+c_7\\
\hline
\end{array}
&~~&
\begin{array}{|c|c|c|c|c|}\hline
\mbox{\tiny DB1}&\mbox{\tiny DB2}&\mbox{\tiny DB3}\\\hline
a_1,b_1,c_1&a_2,b_2,c_2&a_3, b_3, c_3\\
a_2+c_4&a_1+c_8&a_1+c_{12}\\
b_2+c_5&b_1+c_9&b_1+c_{13}\\
a_3+c_6&a_3+c_{10}&a_2+c_{14}\\
b_3+c_7&b_3+c_{11}&b_2+c_{15}\\
a_4+b_4&a_{6}+b_6&a_{8}+b_8\\
a_5+b_5&a_{7}+b_7&a_{9}+b_9\\
a_{6} + b_{6}+c_{16}&a_{4} + b_{4}+c_{20}&a_{4} + b_{4}+c_{24}\\
a_{7} + b_{7}+c_{17}&a_{5} + b_{5}+c_{21}&a_{5} + b_{5}+c_{25}\\
a_{8} + b_{8}+c_{18}&a_{8} + b_{8}+c_{22}&a_{6} + b_{6}+c_{26}\\
a_{9} + b_{9}+c_{19}&a_{9} + b_{9}+c_{23}&a_{7} + b_{7}+c_{27}\\
\hline
\end{array}
\end{array}
\end{eqnarray*}

The structure of the queries is summarized in the following structure matrix. Note again that the structure matrix is made up of sums of place-holders of message bits, no message bit appears more than once, and the assignment of all messages bits to these place-holders is equally likely. 
\begin{eqnarray*}
\begin{array}{r|c|}\cline{2-2}
& \underline{a}\\
& \underline{b}\\
& \underline{c}\\
& \underline{a}+\underline{b}\\
& \underline{a}+\underline{b}\\
& \underline{a}+\underline{c}\\
& \underline{a}+\underline{c}\\
& \underline{b}+\underline{c}\\
& \underline{b}+\underline{c}\\
& \underline{a}+\underline{b}+\underline{c}\\
& \underline{a}+\underline{b}+\underline{c}\\
& \underline{a}+\underline{b}+\underline{c}\\
& \underline{a}+\underline{b}+\underline{c}\\[0.05cm]\cline{2-2}
\end{array}
\end{eqnarray*}

\noindent The  examples illustrated above generalize naturally to arbitrary $N$ and $K$. As we proceed to proofs of privacy and correctness and to calculate the rate  for arbitrary parameters, a more formal algorithmic description will be useful. 

\subsection{Formal Description of Achievable Scheme}
For all $k\in[K]$, define\footnote{Since the number of messages, $K$, can be arbitrary, and we have only $26$ letters in the English alphabet, instead of  $a_i, b_j, c_k$, etc., we now use $u_1(i), u_2(j), u_3(k)$, etc., to represent random permutations of bits from different messages.} vectors $U_k=[u_k(1), u_k(2),\cdots,u_k(N^K)]$. We will use the terminology {\bf $k$-sum} to denote an expression representing the  sum of $k$ distinct  variables, each drawn from a \emph{different} $U_j$ vector, i.e., $u_{j_1}(i_1)+u_{j_2}(i_2)+\cdots+u_{j_k}(i_k)$, where $j_1,j_2,\cdots, j_k\in[K]$ are all \emph{distinct} indices. Furthermore, we will define such a $k$-sum to be of {\bf type} $\{j_1,j_2,\cdots,j_k\}$.

The achievable scheme is comprised of the following elements: 1) a  fixed query set structure, 2) an algorithm to generate the query set as a deterministic function of $\theta$, and 3) a random mapping  from $U_k$ variables to message bits, which will produce the actual queries to be sent to the databases. The random mapping will be privately generated by the user, unknown to the databases. These elements are described next.

\subsubsection{A Fixed Query Set Structure}
For all $\m{DB}\in[N], \theta\in[K]$, let us define `query sets': $Q(\m{DB},\theta)$, which must satisfy the following structural properties. Each $Q(\m{DB},\theta)$ must be the union of $K$ disjoint subsets called ``blocks", that are indexed by $k\in[K]$. Block $k$ must contain only $k$-sums. Note that there are only $\binom{K}{k}$ possible ``\emph{types}" of $k$-sums. Block $k$ must contain all of them. {\color{black} We require that block $k$ contains exactly $(N-1)^{k-1}$ distinct instances of \emph{each type} of $k$-sum. This requirement is chosen following the intuition from the three principles, and as we will prove shortly, it ensures that the resulting scheme is capacity achieving.} Thus, the total number of elements contained in block $k$ must be $\binom{K}{k}(N-1)^{k-1}$, and the total number of elements in each query set must be $|Q(\m{DB},\theta)|=\sum_{k=1}^K\binom{K}{k}(N-1)^{k-1}$. For example, for $N=3, K=3$, as illustrated previously, there are $\binom{3}{1}=3$ types of $1$-sums ($a$, $b$, $c$) and we have $(3-1)^{1-1}=1$ instances of each; there are $\binom{3}{2}=3$ types of $2$-sums ($a+b$, $b+c$, $c+a$) and we have $(3-1)^{2-1}=2$ instances of each; and there is $\binom{3}{3}=1$ type of $3$-sum ($a+b+c$) and we have $(3-1)^{3-1}=4$ instances of it. The query to each database has this structure. Furthermore, no message symbol can appear more than once in a query set for any given database. 



The  structure of Block $k$ of the query $Q(\m{DB},\theta)$, enforced by the constraints described above,  is illustrated in Figure \ref{fig:structure} through an enumeration of all its elements. In the figure, each $\underline{U_j}$ represents a place-holder for a \emph{distinct} element of $U_j$. Note that the structure as represented in Figure \ref{fig:structure} is fixed  regardless of $\theta$ and $\m{DB}$. All query sets must have the same fixed structure.

\allowdisplaybreaks
\begin{figure}[H]
\begin{eqnarray*}
\begin{array}{|l|l|l|c|}\hline
\m{Type No.}&\m{Type of $k$-sum}&\m{Instance No.}&\m{Enumerated elements of } \m{Block } k\\\hline
1.& \{1,2,\cdots,k-2, k-1,k\}&1.&\underline{U_1}+\underline{U_2}+\cdots+\underline{U_{k-2}}+\underline{U_{k-1}}+\underline{U_k}\\
&&2.&\underline{U_1}+\underline{U_2}+\cdots+\underline{U_{k-2}}+\underline{U_{k-1}}+\underline{U_k}\\
&&\vdots&\vdots\\
&&(N-1)^{k-1}.&\underline{U_1}+\underline{U_2}+\cdots+\underline{U_{k-2}}+\underline{U_{k-1}}+\underline{U_k}\\[0.1cm]
\hdashline[1pt/1.5pt]
2.& \{1,2,\cdots,k-2,k-1,k+1\}&1.&\underline{U_1}+\underline{U_2}+\cdots+\underline{U_{k-2}}+\underline{U_{k-1}}+\underline{U_{k+1}}\\
&&2.&\underline{U_1}+\underline{U_2}+\cdots+\underline{U_{k-2}}+\underline{U_{k-1}}+\underline{U_{k+1}}\\
&&\vdots&\vdots\\
&&(N-1)^{k-1}.&\underline{U_1}+\underline{U_2}+\cdots+\underline{U_{k-2}}+\underline{U_{k-1}}+\underline{U_{k+1}}\\[0.1cm]
\hdashline[1pt/1.5pt]
&\vdots&\vdots&\vdots\\
\hdashline[1pt/1.5pt]
i.&\{i_1,i_2,\cdots,i_k\}&1.&\underline{U_{i_1}}+\underline{U_{i_2}}+\cdots+\underline{U_{i_k}}\\
&&2.&\underline{U_{i_1}}+\underline{U_{i_2}}+\cdots+\underline{U_{i_k}}\\
&&\vdots&\vdots\\
&&(N-1)^{k-1}.&\underline{U_{i_1}}+\underline{U_{i_2}}+\cdots+\underline{U_{i_k}}\\[0.1cm]
\hdashline[1pt/1.5pt]
&\vdots&\vdots&\vdots\\
\hdashline[1pt/1.5pt]
\binom{K}{k}.&\{K-k+1,K-k+2,\cdots,K\}&1.&\underline{U_{K-k+1}}+\underline{U_{K-k+2}}+\cdots+\underline{U_{K}}\\
&&2.&\underline{U_{K-k+1}}+\underline{U_{K-k+2}}+\cdots+\underline{U_{K}}\\
&&\vdots&\vdots\\
&&(N-1)^{k-1}.&\underline{U_{K-k+1}}+\underline{U_{K-k+2}}+\cdots+\underline{U_{K}}\\[0.1cm]
\hline
\end{array}
\end{eqnarray*}
\caption{Structure of Block $k$ of $Q(\m{DB},\theta)$. The structure does not depend on $\theta$ or $\m{DB}$. Each $\underline{U}_j$ is a place-holder for a distinct variable from $U_j$.}\label{fig:structure}
\end{figure}

\subsubsection{A Deterministic Algorithm}
Next we present the algorithm which will  produce $Q(\m{DB},\theta)$ for all $\m{DB}\in[N]$ as function of $\theta$ alone. In particular, this algorithm will determine which $U_j$ variable is assigned to each place-holder value in the query structure described earlier. To present the algorithm we need these definitions.

For each $k\in[K]$, let $\new(U_k)$ be a function that, starting with $u_k(1)$,  returns the ``next" variable in $U_k$ each time it is called with   $U_k$ as its argument. So, for example, the following sequence of  calls to this function:  $\new(U_2), \new(U_1),\new(U_1), \new(U_1)+ \new(U_2)$ will produce $u_2(1), u_1(1), u_1(2), u_1(3)+ u_2(2)$ as the output.


Let us partition each block $k$ into two subsets --- a subset $\mathcal{M}$ that contains the $k$-sums which include a variable from $U_\theta$, and a subset  $\mathcal{I}$ which contains all the remaining $k$-sums which contain no symbols from $U_\theta$.\footnote{The nomenclature $\mathcal{M}$ and $\mathcal{I}$ corresponds to `message' and `interference', respectively.}

Using these definitions the algorithm is presented next.
\allowdisplaybreaks
\begin{algorithm}[H]
\caption{Input: $\theta$. Output: Query sets $Q(\m{DB},\theta)$, $\forall \m{DB}\in[N]$}
\label{alg1}
\begin{algorithmic}[1]{}
\STATE{
Initialize: All query sets are initialized as null sets. Also initialize $\m{Block}\leftarrow 1$;} 
\FOR{$\m{DB}=1:N$}
\STATEx{
\begin{eqnarray}Q(\m{DB},\theta,\m{Block},\mathcal{M})&\leftarrow&\{\new(U_\theta)\}\label{al1}\\
 Q(\m{DB},\theta,\m{Block},\mathcal{I})&\leftarrow&\bigcup_{k\in[K], k\neq\theta}\{ \new(U_k)\}\label{al2}
\end{eqnarray}
}\ENDFOR

\FOR[{{\color{blue}Generate each block...}}]{$\m{Block}=2:K$} 
\FOR[{{\color{blue}for each database...}}]{$\m{DB}=1:N$}
\FOR[{{\color{blue}by looking at all `other' databases, and...}}]{\textbf{each} $\m{DB}'=1:N$ \textbf{and} $\m{DB}'\neq \m{DB}$}
\FOR[{{\color{blue} use  the  `$\mathcal{I}$' terms from their previous block...}}]{\textbf{each\footnotemark} $q\in  {Q(\m{DB}',\theta,\m{Block}-1,\mathcal{I})}$}
\STATEx{
 \begin{eqnarray}
 Q(\m{DB},\theta,\m{Block},\mathcal{M})\leftarrow Q(\m{DB},\theta,\m{Block},\mathcal{M})\cup \{\new(U_\theta)+q\}\label{eq:q}
 \end{eqnarray} \COMMENT{{\color{blue} ...to create new $\mathcal{M}$ terms for this block by adding a new $U_\theta$ variable to each term.}}
}\ENDFOR \m{ ($q$)}
 \ENDFOR \m{ ($\m{DB}'$)}
\FOR[{{\color{blue}For  all ``types" that do not include $\theta$...}}]{\textbf{all distinct} $\{i_1,i_2,\cdots,i_{\mbox{\tiny Block}}\}\subset[K]/\{\theta\}$}
\FOR[{{\color{blue}generate exactly $(N-1)^{\m{\tiny Block}-1}$ new instances of each.}}]{$i=1:(N-1)^{\m{\tiny Block}-1}$}
\STATEx{
 $$Q(\m{DB},\theta,\m{Block},\mathcal{I})\leftarrow Q(\m{DB},\theta,\m{Block},\mathcal{I})\cup\{ \new(U_{i_1})+ \new(U_{i_2})+\cdots+ \new(U_{i_{\mbox{\tiny Block}}})\}$$
}
\ENDFOR \m{ ($i$)
}
\ENDFOR \m{ ($\{i_1,i_2,\cdots,i_{\mbox{\tiny Block}}\}$)}
\ENDFOR \m{ (DB)}
\ENDFOR \m{ (Block)}
\FOR{$\m{DB}=1:N$}
\STATE {
$
Q(\m{DB},\theta)\leftarrow\bigcup_{\m{\tiny Block}\in[K]} \big(Q(\m{DB},\theta,\m{Block},\mathcal{I})\cup Q(\m{DB},\theta,\m{Block},\mathcal{M})\big) \label{ni}
$}
\ENDFOR
\end{algorithmic}
\end{algorithm}

\footnotetext{\color{black} For any set $Q$, when accessing its elements in an algorithm (e.g., for all $q\in Q$, do $\ldots$), the output of the algorithm will in general depend on the order in which the elements are accessed. However, for our algorithmic descriptions the order is not important, i.e., any form of ordered access produces an  optimal PIR scheme. By default, a natural lexicographic ordering may be assumed.}

{\color{black}Algorithm \ref{alg1} realizes the 3 principles as follows. The for-loop in steps 5 to 14 ensures database symmetry (principle (1)). The for-loop in steps 10 to 13 ensures message symmetry within one database (principle (2)). Steps 7 to 8 retrieve new desired information using existing side information (principle (3)).}

The proof that the $Q(\m{DB},\theta)$ produced by this algorithm indeed satisfy the query structure described before, is presented in Lemma \ref{lemma:structure}.

\subsubsection{Ordered Representation and  Mapping to Message Bits  to Produce $Q_{\m{\tiny DB}}^{[\theta]}$}
It is useful at this point to have an ordered vector representation of the query structure, as well as the query set $Q(\m{DB},\theta)$. For the query  structure, let us first order the blocks in increasing order of block index. Then within the $k$-th block, $k\in[K]$, arrange the ``types" of $k$-sums by first sorting the indices into $(i_1,i_2,\cdots, i_k)$ such that $i_1<i_2<\cdots<i_k$, and then arranging the $k$-tuples $(i_1,i_2,\cdots, i_k)$ in increasing lexicographic order. For the query set, we have the same arrangement for blocks and types, but then for each given type, we further sort the multiple instances of that type by the $i$ index of the $u_k(i)$ term with the smallest $k$ value in that type. Let $\vec{Q}(\m{DB},\theta)$ denote the ordered representation of $Q(\m{DB},\theta)$. Next we will map the $u_k(i)$ variables to message bits to produce a query vector.

Suppose each message $W_k$, $k\in[K]$, is represented by the vector $W_k=[w_k(1), w_k(2),\cdots, w_k(N^K)]$, where $w_k(i)$ is the binary random variable representing the $i$-th bit of $W_k$. The user privately chooses permutations $\gamma_1, \gamma_2, \cdots, \gamma_K$, uniformly randomly from all possible $(N^K)!$ permutations over the index set $[N^K]$, so that the permutations are independent of each other and of $\theta$. The $U_k$ variables are mapped to the messages $W_k$ through the random permutation $\gamma_k$, $\forall k\in[K]$. Let $\Gamma$ denote an operator that replaces every instance of $u_k(i)$ with $w_k(\gamma_k(i))$, $\forall k\in[K], i\in[N^K]$. For example, $\Gamma(\{u_1(2),u_3(4)+u_5(6)\})=\{w_1(\gamma_1(2)),w_3(\gamma_3(4))+w_5(\gamma_5(6))\}$. This random mapping, applied to $\vec{Q}(\m{DB},\theta)$ produces the actual query vector $Q_{\m{\tiny DB}}^{[\theta]}$ that is  sent to database $\m{DB}$ as
\begin{eqnarray}
Q_{\m{\tiny DB}}^{[\theta]}&=& ``\Gamma\big(\vec{Q}(\m{DB},\theta)\big)"
\end{eqnarray}
We use the double-quotes notation around a random variable to represent the \emph{query} about its realization.  For example, while $w_1(1)$ is a random variable, which may take the value $0$ or $1$, in our notation ``$w_1(1)$" is not random, because it only represents the \emph{question}: ``what is the value of $w_1(1)$?" This is an important distinction, in light of constraints such as (\ref{qwind}) which require that queries must be independent of messages, i.e., message realizations. Note that our queries are indeed independent of message realizations because the queries are generated by the user with no knowledge of message realizations. Also note that the only randomness in $Q_{\m{\tiny DB}}^{[\theta]}$ is because of the $\theta$ and the random permutation $\Gamma$.

\subsection{The Two Examples Revisited}
To illustrate the algorithmic formulation, let us revisit the two examples that were presented previously from an intuitive standpoint.
\subsubsection{Example 1: $N=2, K=2$}
Consider the simplest PIR setting, with $N=2$ databases, and $K=2$ messages with $L = N^K = 4$ bits per message. Instead of our usual notation, i.e., $U_1 = [u_1(1),u_1(2),u_1(3),u_1(4)]$, for this example it will be less cumbersome to use the notation $U_1 = [a_1, a_2, a_3, a_4]$. Similarly, $U_2 = [b_1, b_2, b_3, b_4]$. The query structure and the outputs produced by the algorithm for $\theta=1$ as well as for $\theta=2$ are shown below. The blocks are separated by horizontal lines. Within each block the $\mathcal{I}$ terms are highlighted in red and the $\mathcal{M}$ terms  are in black. Note that there are no terms in $\mathcal{I}$ for the last block (Block $K$), because there are no $K$-sums that do not include the $U_\theta$ variables. 


\begin{eqnarray*}
\begin{array}{rrc}
\m{Query  Structure}&\m{Ordered Output of Algorithm  \ref{alg1} for $\theta=1$}&\m{Ordered Output of Algorithm  \ref{alg1} for $\theta=2$}\\
\begin{array}{r|c|}\cline{2-2}
&{\rule{0pt}{1.2em}\small \vec{Q}(\m{DB},\theta)}\\
\cline{2-2}
\mbox{\tiny Block 1}&\underline{U_1}\\
&\underline{U_2}\\[0.05cm]\cline{2-2}
\mbox{\tiny Block 2}&\underline{U_1}+\underline{U_2}\\[0.05cm]
\cline{2-2}
\end{array}
&~
\begin{array}{r|c|r|c|c|}\cline{2-2}\cline{4-4}
&\rule{0pt}{1.2em}\vec{Q}(\m{DB}1,\theta=1)  & &\rule{0pt}{1.2em}\vec{Q}(\m{DB}2,\theta=1)\\\cline{2-2}\cline{4-4}
\mbox{\tiny }& a_1& \mbox{\tiny }& a_2 \\
 \mbox{\tiny }&{\color{red}b_1}& \mbox{\tiny }& {\color{red}b_2} \\\cline{2-2}\cline{4-4}
 \mbox{\tiny }&a_3+b_2 & \mbox{\tiny }& a_4+b_1 \\\cline{2-2}\cline{4-4}
\end{array}
&
\begin{array}{l|c|c|c|c|}\cline{2-2}\cline{4-4}
&\rule{0pt}{1.2em}\vec{Q}(\m{DB}1,\theta=2)  & &\rule{0pt}{1.2em}\vec{Q}(\m{DB}2,\theta=2)\\\cline{2-2}\cline{4-4}
\mbox{\tiny } &{\color{red}a_1}& \mbox{\tiny }& {\color{red}a_2} \\
\mbox{\tiny }& b_1& \mbox{\tiny }& b_2 \\\cline{2-2}\cline{4-4}
 \mbox{\tiny }&a_2+b_3  &\mbox{\tiny }& a_1+b_4 \\\cline{2-2}\cline{4-4}
\end{array}
\end{array}
\end{eqnarray*}


To verify that the scheme is correct, note that whether $\theta=1$ or $\theta=2$, every desired bit is either downloaded directly (block 1) or appears with known side information that is available from the other database.  To see why privacy holds, recall that the queries are ultimately presented to the database in terms of the message variables and the mapping from $U_k$ to $W_k$ is uniformly random and independent of $\theta$. So, consider an arbitrary realization of the query with (distinct) message bits $w_1(i_1),w_2(i_2)$ from $W_1$ and $w_2(j_1),w_2(j_2)$ from $W_2$. 
\begin{eqnarray}
\begin{array}{|c|}\hline
\Gamma(\rule{0pt}{1.2em}\vec{Q}(\m{DB},\theta))\\
\hline
w_1(i_1)\\
w_2(j_1)\\
\hline
w_1(i_2)+w_2(j_2)\\
\hline
\end{array}
\end{eqnarray}
Given this query, the probability that it was generated for $\theta=1$ is $((\frac{1}{4})(\frac{1}{3}))^2=\frac{1}{144}$, which is the same as the probability that it was generated for $\theta=2$. Thus, the query provides the database no information about $\theta$, and the scheme is private. This  argument is presented in  detail and generalized to arbitrary $K$ and $N$ in Lemma \ref{lemma:private}.
Finally, consider the rate of this scheme. The total number of downloaded bits is $6$, and the number of desired bits  downloaded is $4$, so the rate of this scheme is $4/6=2/3$ which matches the capacity for this case. 

\subsubsection{Example 2: $N=3, K=3$}

The second example is when $K = 3$, $N = 3$. In this case, both messages have $L = N^K = 27$ bits.  $U_1=[a_1, a_2, \cdots, a_{27}], U_2 = [b_1, b_2, \cdots,, b_{27}], U_3 = [c_1, c_2,\cdots, c_{27}]$. The query structure and the output of the algorithm for $\theta=1$ are shown below.
\begin{eqnarray*}
\begin{array}{rcc}
\m{Query  Strucure}&~~&\m{Ordered Output of Algorithm \ref{alg1} for $\theta=1$}\\
\begin{array}{r|c|}\cline{2-2}
& \rule{0pt}{1.2em}\vec{Q}(\m{DB},\theta)\\\cline{2-2}
\m{\tiny Block 1} & \underline{U_1}\\
& \underline{U_2}\\
& \underline{U_3}\\[0.05cm]
\cline{2-2}
\m{\tiny Block 2} & \underline{U_1}+\underline{U_2}\\
& \underline{U_1}+\underline{U_2}\\
& \underline{U_1}+\underline{U_3}\\
& \underline{U_1}+\underline{U_3}\\
& \underline{U_2}+\underline{U_3}\\
& \underline{U_2}+\underline{U_3}\\[0.05cm]\cline{2-2}
\m{\tiny Block 3} & \underline{U_1}+\underline{U_2}+\underline{U_3}\\
& \underline{U_1}+\underline{U_2}+\underline{U_3}\\
& \underline{U_1}+\underline{U_2}+\underline{U_3}\\
& \underline{U_1}+\underline{U_2}+\underline{U_3}\\[0.05cm]\cline{2-2}
\end{array}
&
&
\begin{array}{rl|c|cc|c|cc|c|}\cline{3-3}\cline{6-6}\cline{9-9}
&&\rule{0pt}{1.2em}\vec{Q}(\m{DB}1,\theta=1)  & &&\rule{0pt}{1.2em}\vec{Q}(\m{DB}2,\theta=1) &&& \rule{0pt}{1.2em}\vec{Q}(\m{DB}3,\theta=1) \\\cline{3-3}\cline{6-6}\cline{9-9}
&\mbox{\tiny }& a_1& &\mbox{\tiny }& a_2 &&\mbox{\tiny }& a_3\\
&\mbox{\tiny }& {\color{red}b_1}& &\mbox{\tiny }& {\color{red}b_2} &&\mbox{\tiny }& {\color{red}b_3}\\
&\mbox{\tiny }& {\color{red}c_1}& &\mbox{\tiny }& {\color{red}c_2} &&\mbox{\tiny }& {\color{red}c_3}
\\\cline{3-3}\cline{6-6} \cline{9-9}
&\mbox{\tiny }& {\color{black}a_4+b_2}& &\mbox{\tiny }& {\color{black}a_8+b_1} &&\mbox{\tiny }& {\color{black}a_{12}+b_1}\\
&\mbox{\tiny }& {\color{black}a_6+b_3}& &\mbox{\tiny }& {\color{black}a_{10}+b_3} &&\mbox{\tiny }& {\color{black}a_{14}+b_2}\\
&\mbox{\tiny }& {\color{black}a_5+c_2}& &\mbox{\tiny }& {\color{black}a_{9}+c_1} &&\mbox{\tiny }& {\color{black}a_{13}+c_1}\\
&\mbox{\tiny }& {\color{black}a_7+c_3}& &\mbox{\tiny }& {\color{black}a_{11}+c_3} &&\mbox{\tiny }& {\color{black}a_{15}+c_2}\\
&\mbox{\tiny }& {\color{red}b_4+c_4}& &\mbox{\tiny }& {\color{red}b_{6}+c_6} &&\mbox{\tiny }& {\color{red}b_{8}+c_8}\\
&\mbox{\tiny }& {\color{red}b_5+c_5}& &\mbox{\tiny }& {\color{red}b_{7}+c_7} &&\mbox{\tiny }& {\color{red}b_{9}+c_9}\\\cline{3-3} \cline{6-6} \cline{9-9}
&\mbox{\tiny }& {\color{black}a_{16}+b_6+c_6}& &\mbox{\tiny }& {\color{black}a_{20}+b_4+c_4} &&\mbox{\tiny }& {\color{black}a_{24}+b_4+c_4}\\
&\mbox{\tiny }& {\color{black}a_{17}+b_7+c_7}& &\mbox{\tiny }& {\color{black}a_{21}+b_5+c_5} &&\mbox{\tiny }& {\color{black}a_{25}+b_5+c_5}\\
&\mbox{\tiny }& {\color{black}a_{18}+b_8+c_8}& &\mbox{\tiny }& {\color{black}a_{22}+b_8+c_8} &&\mbox{\tiny }& {\color{black}a_{26}+b_6+c_6}\\
&\mbox{\tiny }& {\color{black}a_{19}+b_9+c_9}& &\mbox{\tiny }& {\color{black}a_{23}+b_9+c_9} &&\mbox{\tiny }& {\color{black}a_{27}+b_7+c_7}\\\cline{3-3} \cline{6-6} \cline{9-9}
\end{array}
\end{array}
\end{eqnarray*}

The output of Algorithm \ref{alg1}, for $\theta=2$, is shown next.
\begin{eqnarray*}
{\small
\begin{array}{rl|c|cc|c|cc|c|}\cline{3-3}\cline{6-6}\cline{9-9}
&&\rule{0pt}{1.2em}\vec{Q}(\m{DB}1,\theta=2)  &&&\rule{0pt}{1.2em}\vec{Q}(\m{DB}2,\theta=2) &&& \rule{0pt}{1.2em}\vec{Q}(\m{DB}3,\theta=2) \\\cline{3-3}\cline{6-6}\cline{9-9}
&\mbox{\tiny }&{\color{red}a_1}&&\mbox{\tiny }& {\color{red}a_2} &&\mbox{\tiny }& {\color{red}a_3}\\
&\mbox{\tiny }&{\color{black}b_1}  &&\mbox{\tiny }& {\color{black}b_2}  &&\mbox{\tiny }& {\color{black}b_3} \\
&\mbox{\tiny }&{\color{red}c_1}  &&\mbox{\tiny }&  {\color{red}c_2}  &&\mbox{\tiny }&  {\color{red}c_3} \\\cline{3-3}\cline{6-6} \cline{9-9}
&\mbox{\tiny }&  a_2+b_4   &&\mbox{\tiny }& a_1+b_8   &&\mbox{\tiny }& a_{1}+b_{12}  \\
&\mbox{\tiny }&  a_3+b_6  &&\mbox{\tiny }& a_{3}+b_{10}  &&\mbox{\tiny }& a_{3}+b_{14} \\
&\mbox{\tiny }&  {\color{red}a_4+c_4}  & &\mbox{\tiny }& {\color{red}a_6+c_6}  &&\mbox{\tiny }&  {\color{red}a_{8}+c_8}  \\
&\mbox{\tiny }&   {\color{red}a_5+c_5} & &\mbox{\tiny }& {\color{red} a_{7}+c_7} &&\mbox{\tiny }& {\color{red}a_{9}+c_9} \\
&\mbox{\tiny }& {\color{black}b_5+c_2} & &\mbox{\tiny }& {\color{black}b_9+c_1}  &&\mbox{\tiny }& {\color{black}b_{13}+c_1}  \\
&\mbox{\tiny }& {\color{black}b_7+c_3} & &\mbox{\tiny }& {\color{black}b_{11}+c_3} &&\mbox{\tiny }&  {\color{black}b_{15}+c_3} \\
\cline{3-3} \cline{6-6} \cline{9-9}
&\mbox{\tiny }&   a_{6}+b_{16}+c_6  & &\mbox{\tiny }& a_{4}+b_{20}+c_4 &&\mbox{\tiny }& a_{4}+b_{24}+c_4  \\
&\mbox{\tiny }&  a_{7}+b_{17}+c_7& &\mbox{\tiny }&a_{5}+b_{21}+c_5 &&\mbox{\tiny }& a_{5}+b_{25}+c_5  \\
&\mbox{\tiny }&   a_{8}+b_{18}+c_8   & &\mbox{\tiny }& a_{8}+b_{22}+c_8 &&\mbox{\tiny }& a_{6}+b_{26}+c_6 \\
&\mbox{\tiny }&   a_{9}+b_{19}+c_9  & &\mbox{\tiny }& a_{9}+b_{23}+c_9 &&\mbox{\tiny }& a_{7}+b_{27}+c_7  \\
\cline{3-3} \cline{6-6} \cline{9-9}
\end{array}}
\end{eqnarray*}
The output of Algorithm \ref{alg1}, for $\theta=3$, is shown next.
\begin{eqnarray*}
{\small
\begin{array}{rl|c|cc|c|cc|c|}\cline{3-3}\cline{6-6}\cline{9-9}
&&\rule{0pt}{1.2em}\vec{Q}(\m{DB}1,\theta=3)  & &&\rule{0pt}{1.2em}\vec{Q}(\m{DB}2,\theta=3) &&& \rule{0pt}{1.2em}\vec{Q}(\m{DB}3,\theta=3) \\\cline{3-3}\cline{6-6}\cline{9-9}
&\mbox{\tiny }& {\color{red} a_1}& &\mbox{\tiny }& {\color{red} a_2} &&\mbox{\tiny }& {\color{red} a_3}\\
&\mbox{\tiny }& {\color{red}b_1}& &\mbox{\tiny }& {\color{red}b_2} &&\mbox{\tiny }& {\color{red}b_3}\\
&\mbox{\tiny }& {\color{black}c_1}& &\mbox{\tiny }& {\color{black}c_2} &&\mbox{\tiny }& {\color{black}c_3}
\\\cline{3-3}\cline{6-6} \cline{9-9}
&\mbox{\tiny }& {\color{red}a_4+b_4}& &\mbox{\tiny }& {\color{red}a_6+b_6} &&\mbox{\tiny }& {\color{red}a_{8}+b_8}\\
&\mbox{\tiny }& {\color{red}a_5+b_5}& &\mbox{\tiny }& {\color{red}a_{7}+b_7} &&\mbox{\tiny }& {\color{red}a_{9}+b_9}\\
&\mbox{\tiny }& {\color{black}a_2+c_4}& &\mbox{\tiny }& {\color{black}a_{1}+c_{8}} &&\mbox{\tiny }& {\color{black}a_{1}+c_{12}}\\
&\mbox{\tiny }& {\color{black}a_3+c_6}& &\mbox{\tiny }& {\color{black}a_{3}+c_{10}} &&\mbox{\tiny }& {\color{black}a_{2}+c_{14}}\\
&\mbox{\tiny }& {\color{black}b_2+c_5}& &\mbox{\tiny }& {\color{black}b_{1}+c_9} &&\mbox{\tiny }& {\color{black}b_{1}+c_{13}}\\
&\mbox{\tiny }& {\color{black}b_3+c_7}& &\mbox{\tiny }& {\color{black}b_{3}+c_{11}} &&\mbox{\tiny }& {\color{black}b_{2}+c_{15}}\\\cline{3-3} \cline{6-6} \cline{9-9}
&\mbox{\tiny }& {\color{black}a_{6}+b_6+c_{16}}& &\mbox{\tiny }& {\color{black}a_{4}+b_4+c_{20}} &&\mbox{\tiny }& {\color{black}a_{4}+b_4+c_{24}}\\
&\mbox{\tiny }& {\color{black}a_{7}+b_7+c_{17}}& &\mbox{\tiny }& {\color{black}a_{5}+b_5+c_{21}} &&\mbox{\tiny }& {\color{black}a_{5}+b_5+c_{25}}\\
&\mbox{\tiny }& {\color{black}a_{8}+b_8+c_{18}}& &\mbox{\tiny }& {\color{black}a_{8}+b_8+c_{22}} &&\mbox{\tiny }& {\color{black}a_{6}+b_6+c_{26}}\\
&\mbox{\tiny }& {\color{black}a_{9}+b_9+c_{19}}& &\mbox{\tiny }& {\color{black}a_{9}+b_9+c_{23}} &&\mbox{\tiny }& {\color{black}a_{7}+b_7+c_{27}}\\\cline{3-3} \cline{6-6} \cline{9-9}
\end{array}}
\end{eqnarray*}
Note that this construction retrieves $27$ desired message bits out of a total of $39$ downloaded bits, so its rate is $27/39=9/13$, which matches the capacity for this case.

\subsection{Proof of Correctness, Privacy and Achieving Capacity}
The following lemma confirms that the query set produced by the algorithm satisfies the required structural properties.

\begin{lemma}\label{lemma:structure}$(${\bf Structure of} $Q(\m{DB},\theta))$ For any $\theta\in[K]$ and for any $\m{DB}\in[N]$,  the $Q(\m{DB},\theta)$ produced by Algorithm 1 satisfies the following properties.
\begin{enumerate}
\item For all $k\in[K]$, block $k$ contains  exactly $(N-1)^{k-1}$ instances of $k$-sums of each possible type.
\item No $u_k(i), i \in [N^K]$ variable appears more than once within $Q(\m{DB},\theta)$ for any given $\m{DB}$.
\item Exactly  $N^{K-1}$ variables for each $U_k$, $k\in[K]$, appear in the query set $Q(\m{DB},\theta)$.
\item The size of $Q(\m{DB},\theta)$ is $N^{K-1} + \frac{1}{N-1}(N^{K-1}-1)$.
\end{enumerate}
\end{lemma}

\proof 
\begin{enumerate}
\item Fix any arbitrary $N$. The proof is based on induction on the claim $S(k)$, defined as follows. \\
$S(k):$ ``{\it Block $k$ contains exactly $(N-1)^{k-1}$ instances of $k$-sums of all possible types.}"

 The basis step is when $k = 1$. This step is easily verified, because a $1$-sum is simply one variable, of which there are $K$ possible types, and from (\ref{al1}), (\ref{al2}) in Algorithm \ref{alg1}, we note that the first block always consists of one variable of each vector $U_k, k \in [K]$. 

We next proceed to the inductive step. Suppose $S(k)$ is true. Then we wish to prove that $S(k+1)$ must be true as well. Here we  have $\m{Block}=k+1$. First, consider $(k+1)$-sums of type $\{i_1, i_2, \cdots, i_{k+1}\}\subset[K]/\{\theta\}$ where none of the indices is $\theta$. These belong in $Q(\m{DB}, \theta,k+1,\mathcal{I})$, and from line 11 of the algorithm it is verified that exactly $(N-1)^{\m{\tiny Block}-1}=(N-1)^k$ instances are generated of this type. Next, consider the $(k+1)$-sums of type $\{i_1, i_2, \cdots, i_{k},\theta\}$ where one of the indices is $\theta$. These belong to $Q(\m{DB}, \theta,k+1,\mathcal{M})$ and are obtained by adding $\new(U_\theta)$ to each of the $k$-sums of type $\{i_1, i_2, \cdots, i_{k}\}$ that belong to $Q(\m{DB}',\theta,k,\mathcal{I})$ for all $\m{DB}'\neq\m{DB}$. Therefore, the number of instances of $(k+1)$-sums of type  $\{i_1, i_2, \cdots, i_{k},\theta\}$ in $Q(\m{DB}, \theta,k+1,\mathcal{M})$ must be equal to the product of the number of `other' databases $\m{DB}'$, which is equal to $N-1$, and the number of instances of type $\{i_1, i_2, \cdots, i_{k}\}$ in each database $\m{DB}'$, which is equal to $(N-1)^{k-1}$ because $S(k)$ is assumed to be true as the induction hypothesis. $(N-1)\times(N-1)^{k-1}=(N-1)^k$, and thus, we have shown that $S(k+1)$ is true, completing the proof by induction.

%


\item From (\ref{al1}),(\ref{eq:q}), we see that for each block, the desired variables, i.e., the $U_\theta$ variables appear only through the  $\new(U_\theta)$ function so that each of them only appears once. For the non-desired variables $U_k, k\neq\theta$, we see that the only time that they do not appear through the $\new(U_k)$ function is when they enter through  $q$ in (\ref{eq:q}).  However, from (\ref{eq:q}) we see that these variables come from the $\mathcal{I}$ part of the previous block of other databases, where each of them was only introduced once through a  $\new(U_k)$ function. Moreover, each term from the $\mathcal{I}$ part of the previous block of other databases is used exactly once. Therefore, these $U_k$ variables also appear no more than once in the query set of a given database. 

\item Since we have shown that no variable appears more than once, we only need to count the number of times each vector $U_k, k \in [K]$ is invoked within $Q(\m{DB},\theta)$. Consider any particular vector, say $U_j$.  The number of possible types of $k$-sums that include index $j$ is  $\binom{K-1}{k-1}$. As we have also shown, the $k$-th block contains $(N-1)^{k-1}$ instances of $k$-sums of each type. Therefore, the number of instances of vector $U_j$ in block $k$ is $(N-1)^{k-1}\binom{K-1}{k-1}$. Summing over all $K$ blocks within $Q(\m{DB}, \theta)$ we find
\begin{eqnarray}
 \sum_{k=1}^K  (N-1)^{k-1} \binom{K-1}{k-1} &=& (N-1+1)^{K-1} = N^{K-1}~~~~\mbox{(Binomial Identity)} \label{eq:binid}
\end{eqnarray}

%

\item The $k$-th block of $Q(\m{DB},\theta)$ contains  $(N-1)^{k-1}$ instances of $k$-sums of each possible type, and there are $\binom{K}{k}$ possible types of $k$-sums. Therefore, the cardinality of  $Q(\m{DB},\theta)$ is
\begin{eqnarray}
|Q(\m{DB},\theta)|&=& \sum_{k=1}^K (N-1)^{k-1}  \binom{K}{k} \\
& \overset{}{=}& \sum_{k=1}^K (N-1)^{k-1} \left[ \binom{K-1}{k} + \binom{K-1}{k-1}\right] \\
&\overset{(\ref{eq:binid})}{=}& N^{K-1} +  \sum_{k=1}^{K-1} (N-1)^{k-1}  \binom{K-1}{k} \\
&\overset{}{=}& N^{K-1} +  \frac{1}{N-1} \sum_{k=1}^{K-1} (N-1)^{k}  \binom{K-1}{k}\\
&\overset{}{=}& N^{K-1} +  \frac{1}{N-1} \left[ \sum_{k=0}^{K-1} (N-1)^{k}  \binom{K-1}{k} - 1\right] \\
&\overset{}{=}& N^{K-1} +  \frac{1}{N-1} (N^{K-1} - 1)
\end{eqnarray}

%
\end{enumerate}
\hfill\QED

We are now ready to prove that the achievable scheme is correct, private and achieves the capacity, in the following two lemmas.

\begin{lemma}
The scheme described in Algorithm \ref{alg1} is correct and the rate achieved is $(1 + 1/N + \cdots + 1/N^{K-1})^{-1}$, which matches the capacity.
\end{lemma}
\proof 
The scheme is correct, i.e., all  desired  variables, $U_\theta$,  are decodable (with zero error probability), because either they appear with no interference (the first block) or they appear with interference $q$ that is also downloaded separately from another database $\m{DB}'$ so it can be subtracted. From Lemma \ref{lemma:structure}  we know that there are $N^{K-1}$ desired bit-variables in each $Q(\m{DB},\theta)$. Note that desired variables always appear through $\new(U_\theta)$, so they do not repeat across databases. Thus, the total number of desired bits that are retrieved is $N\times N^{K-1}=N^K$.


We next compute the rate. The total number of desired bits retrieved is $N^K$, and the total number of downloaded bits from all databases is $N\times|Q(\m{DB},\theta)|$ in every case. Therefore, the rate,
\begin{eqnarray}
R &=& \frac{N^K}{N\times|Q(\m{DB},\theta)|} \\
&=& \frac{N^K}{N[N^{K-1} +  \frac{1}{N-1} (N^{K-1} - 1)]} \\
&=& \left(\frac{N^{K-1} + \frac{1}{N-1} (N^{K-1} - 1)}{N^{K-1}} \right)^{-1} = \left(1 + \frac{ \frac{1}{N-1} (N^{K-1} - 1)}{N^{K-1}} \right)^{-1}\\
&=& \left(1 + \frac{ \frac{1}{N} (1 - \frac{1}{N^{K-1}})}{1 - \frac{1}{N}} \right)^{-1} = \left( 1+\frac{1}{N} + \cdots + \frac{1}{N^{K-1}} \right)^{-1}
\end{eqnarray}

\hfill\QED

\begin{lemma}
The scheme described in Algorithm \ref{alg1} is private.\label{lemma:private}
\end{lemma}
\proof 
The intuition is quite straightforward. Regardless of $\theta$, every realization of the query vector that fits the query structure is equally likely because of the uniformly random permutation $\Gamma$. To formalize this intuition, let us  calculate the probability of an arbitrary query realization.

For any $\m{DB}\in[N], \theta\in[K]$, consider the ordered query vector representation $\vec{Q}(\m{DB},\theta)$. 
For each $U_k$, $k\in[K]$,  denote the order in which these symbols appear in $\vec{Q}(\m{DB},\theta)$, as $\vec{u}_k(\m{DB},\theta)=[u_k(i_{k,\m{\tiny DB},\theta,1}),u_k(i_{k,\m{\tiny DB},\theta,2}),\cdots,u_k(i_{k,\m{\tiny DB},\theta,N^{K-1}})]$. Since the ordered query structure is already fixed regardless of $\theta$ and $\m{DB}$, and no variable occurs more than once, $\vec{Q}(\m{DB},\theta)$ is completely determined by $(\vec{u}_1(\m{DB},\theta), \vec{u}_2(\m{DB},\theta),\cdots, \vec{u}_K(\m{DB},\theta))$. Similarly, for each $k\in[K]$, denote an \emph{arbitrary} $N^{K-1}$-tuple of bits from message $W_k$ by $\vec{w}_k=[w_k(i'_{k_1}),w_k(i'_{k_2}),\cdots,w_k(i'_{k_{N^{K-1}}})]$. Recall that $u_k(i)=w_k(\gamma_k(i))$, $\forall k\in[K], i\in[N^K]$, and $\gamma_1,\gamma_2,\cdots,\gamma_K$ are uniform permutations chosen independently of each other and also independently of $\theta$. Therefore, for all $(\vec{w}_1,\vec{w}_2,\cdots,\vec{w}_K)$, we have
\begin{eqnarray}
\lefteqn{\m{Prob}\Big(\Gamma\left(\vec{u}_1(\m{DB},\theta),\vec{u}_2(\m{DB},\theta),\cdots,\vec{u}_K(\m{DB},\theta)\right)=\left(\vec{w}_1,\vec{w}_2,\cdots,\vec{w}_K\right)\Big)}\nonumber\\
&=&\prod_{k=1}^K\m{Prob}\Big(\Gamma(\vec{u}_k(\m{DB},\theta))=\vec{w}_k\Big)\\
&=&\left(\left(\frac{1}{N^K}\right)\left(\frac{1}{N^K-1}\right)\cdots\left(\frac{1}{N^K-N^{K-1}+1}\right)\right)^K \label{eq:queryp}
\end{eqnarray}
which does not depend on $\theta$. Thus, the distribution of  $\vec{Q}(\m{DB},\theta)$ does not depend on $\theta$. Since  $Q_{\m{\tiny DB}}^{[\theta]}$ is a function of $\vec{Q}(\m{DB},\theta)$,  $Q_{\m{\tiny DB}}^{[\theta]}$ must be independent  of $\theta$ as well. Next, we show that  privacy requirement (\ref{privacy}) must be satisfied. 
\begin{eqnarray}
I(\theta; Q_{\m{\tiny DB}}^{[\theta]}, A_{\m{\tiny DB}}^{[\theta]}, W_{1:K}) &=& I(\theta;Q_{\m{\tiny DB}}^{[\theta]}) + I(\theta; W_{1:K}|Q_{\m{\tiny DB}}^{[\theta]}) +  I(\theta; A_{\m{\tiny DB}}^{[\theta]}|W_{1:K}, Q_{\m{\tiny DB}}^{[\theta]}) \\
&=& 0 + 0 + 0 = 0 \label{eq:o}
\end{eqnarray}
where $I(\theta;Q_{\m{\tiny DB}}^{[\theta]}) = 0$ because we have already proved that $Q_{\m{\tiny DB}}^{[\theta]}$ is independent of $\theta$,  $I(\theta; W_{1:K}|Q_{\m{\tiny DB}}^{[\theta]}) = 0$ because the desired message index and the query are generated privately by the user with no knowledge of the messages, and $I(\theta; A_{\m{\tiny DB}}^{[\theta]}|W_{1:K}, Q_{\m{\tiny DB}}^{[\theta]}) = 0$ because the answer is deterministic function of the query and messages. Therefore, all information available to database DB ($Q_{\m{\tiny DB}}^{[\theta]}, A_{\m{\tiny DB}}^{[\theta]}, W_1, \cdots, W_K$) is independent of $\theta$ and the scheme is private.\hfill\QED

{\it Remark: From the proofs of privacy and correctness, note that the key is the query structure and the random mapping, $\Gamma$, of message bits to the query structure. In particular, no assumption is required on the statistics of the messages themselves. So the scheme works and a rate equal to $C_o$ remains achievable even if the messages are not independent, although it may no longer be the capacity for this setting. For example, if $N=K=2$ and the two messages are identical, $W_1=W_2$, then clearly the capacity is $1$, which is higher than $C_o=2/3$. The independence of the messages is, however, needed for the converse.}

We end this section with a lemma that highlights a curious property of our capacity achieving PIR scheme -- that if the  scheme is projected onto any subset of messages by eliminating the remaining messages, it also achieves the PIR capacity for that subset of messages.

\begin{lemma}
Given a capacity achieving scheme generated by Algorithm \ref{alg1} for $K$ messages, if we set $\Delta, 1 \leq \Delta \leq K-1$ messages to be null, then the scheme achieves the capacity for the remaining $K - \Delta$ messages.
\end{lemma}

\proof We first prove that the scheme is correct after eliminating messages.  This is easy to see as eliminating messages does not hurt (influence) the decoding procedure. 
Note that the eliminated messages can not include the desired one. We next prove that the scheme is also private. This is also easy to see as the permutations of the messages are independent, so that after eliminating messages, the bits of the remaining messages still distribute identically, no matter which message is desired. We finally compute the rate and show that the scheme achieves the capacity for the remaining messages. Note that the total number of desired bits does not change, i.e., it is still $N^K$. The total number of downloaded equations decreases, as $\Delta$ messages are set to 0. In particular, the following number of equations becomes 0.
\begin{eqnarray}
&& N \sum_{k = 1}^\Delta \binom{\Delta}{k} (N-1)^{k-1} \\
&=& N \frac{1}{N-1} \left[ \sum_{k = 0}^\Delta \binom{\Delta}{k} (N-1)^{k} - 1\right] \\
&=& N \frac{1}{N-1} (N^\Delta - 1)
\end{eqnarray} 
Subtracting above from $N |Q(\m{DB},\theta)|$, we have the total number of downloaded equations. Therefore, the rate achieved is
\begin{eqnarray}
R &=& \frac{N^K}{N |Q(\m{DB},\theta)| - N \frac{1}{N-1} (N^\Delta - 1) } \\
&=& \frac{N^K}{N[N^{K-1} +  \frac{1}{N-1} (N^{K-1} - 1) -  \frac{1}{N-1} (N^\Delta - 1)]} \\
&=& \left(\frac{N^{K-1} + \frac{1}{N-1} (N^{K-1} - N^\Delta)}{N^{K-1}} \right)^{-1} = \left(1 + \frac{ \frac{1}{N-1} (N^{K-1} - N^\Delta)}{N^{K-1}} \right)^{-1}\\
&=& \left(1 + \frac{ \frac{1}{N} (1 - \frac{1}{N^{K-\Delta-1}})}{1 - \frac{1}{N}} \right)^{-1} = \left( 1+\frac{1}{N} + \cdots + \frac{1}{N^{K-\Delta - 1}} \right)^{-1}
\end{eqnarray}
which matches the capacity.

\hfill\QED

\section{Theorem \ref{thm:download}: Converse}\label{sec:con}
Note that the converse is proved for arbitrary $L$, i.e., we no longer assume that $L=N^K$. 
Let us start with two useful lemmas.
{\color{black} Note that in the proofs, the relevant equations needed to justify each step are specified by the equation numbers set on top of the (in)equality symbols.}
\begin{lemma}\label{lemma:int}
$I(W_{2:K}; Q_{1:N}^{[1]}, A_{1:N}^{[1]} | W_1) \leq L(1/R - 1 + o(L))$
\end{lemma}
\proof
\begin{eqnarray}
&& I(W_{2:K}; Q_{1:N}^{[1]}, A_{1:N}^{[1]} | W_1) \notag\\
&\overset{(\ref{h1})}{=}& I(W_{2:K}; Q_{1:N}^{[1]}, A_{1:N}^{[1]}, W_1) \\
&=& I(W_{2:K}; Q_{1:N}^{[1]}, A_{1:N}^{[1]}) + I(W_{2:K}; W_1 | Q_{1:N}^{[1]}, A_{1:N}^{[1]})\\
&\overset{(\ref{corr})}{=}& I(W_{2:K}; Q_{1:N}^{[1]}, A_{1:N}^{[1]}) + o(L)L \\
&\overset{(\ref{qwind})}{=}& I(W_{2:K}; A_{1:N}^{[1]} |Q_{1:N}^{[1]}) + o(L)L \\
&\overset{}{=}& H(A_{1:N}^{[1]} |Q_{1:N}^{[1]})  - H(A_{1:N}^{[1]} |Q_{1:N}^{[1]}, W_{2:K}) + o(L)L \\
&\overset{}{\leq}& D - H(W_1, A_{1:N}^{[1]} |Q_{1:N}^{[1]}, W_{2:K}) + H(W_1| A_{1:N}^{[1]},Q_{1:N}^{[1]}, W_{2:K}) + o(L)L \\
&\overset{(\ref{eta_def})(\ref{ansdet})(\ref{corr})}{=}& L/R - H(W_1 |Q_{1:N}^{[1]}, W_{2:K}) + o(L)L \\
&\overset{(\ref{qwind})(\ref{h1})(\ref{h2})}{=}& L/R - L + o(L)L = L(1/R - 1 + o(L))
\end{eqnarray} 

\hfill\QED

\begin{lemma}\label{lemma:recursive}
For all $k \in \{2,\cdots,K\}$, 
\begin{eqnarray}
I(W_{k:K}; Q_{1:N}^{[k-1]}, A_{1:N}^{[k-1]} | W_{1:k-1}) &\geq& \frac{1}{N} I(W_{k+1:K}; Q_{1:N}^{[k]}, A_{1:N}^{[k]} | W_{1:k}) + \frac{L(1-o(L))}{N}.
\end{eqnarray}
\end{lemma}
\proof
\begin{eqnarray}
&& N I(W_{k:K}; Q_{1:N}^{[k-1]}, A_{1:N}^{[k-1]} | W_{1:k-1}) \notag\\
&\geq& \sum_{n=1}^N  I(W_{k:K}; Q_n^{[k-1]}, A_n^{[k-1]}|W_{1:k-1}) \\
&\overset{(\ref{privacy})}{=}&  \sum_{n=1}^N  I(W_{k:K}; Q_n^{[{\color{blue} k}]}, A_n^{[{\color{blue} k}]}|W_{1:k-1}) \\
&\geq& \sum_{n=1}^N I(W_{k:K}; A_n^{[k]} | W_{1:k-1}, Q_n^{[k]})\\
&\overset{(\ref{ansdet})}{=}& \sum_{n=1}^N H(A_n^{[k]} | W_{1:k-1}, Q_n^{[k]})\\
&\overset{}{\geq}& \sum_{n=1}^N H(A_n^{[k]} | W_{1:k-1}, Q_{1:N}^{[k]}, A_{1:n-1}^{[k]} )\\
&\overset{(\ref{ansdet})}{=}& \sum_{n=1}^N I(W_{k:K};A_n^{[k]} | W_{1:k-1}, Q_{1:N}^{[k]}, A_{1:n-1}^{[k]})\\
&\overset{}{=}&  I(W_{k:K};A_{1:N}^{[k]} | W_{1:k-1},Q_{1:N}^{[k]})\\
&\overset{(\ref{qwind})(\ref{h1})}{=}&  I(W_{k:K};Q_{1:N}^{[k]}, A_{1:N}^{[k]} | W_{1:k-1})\\
&\overset{(\ref{corr})}{=}&  I(W_{k:K};W_k, Q_{1:N}^{[k]}, A_{1:N}^{[k]} | W_{1:k-1}) -  o(L)L \\
&\overset{}{=}&  I(W_{k:K};W_k | W_{1:k-1}) + I(W_{k:K};Q_{1:N}^{[k]}, A_{1:N}^{[k]} | W_{1:k}) - o(L)L \\
&\overset{(\ref{h1})(\ref{h2})}{=}& L + I(W_{k:K};Q_{1:N}^{[k]}, A_{1:N}^{[k]} | W_{1:k}) -  o(L)L \\
&\overset{}{=}& I(W_{k+1:K};Q_{1:N}^{[k]}, A_{1:N}^{[k]} | W_{1:k}) + L(1-o(L)
\end{eqnarray} 

\hfill\QED

With these lemmas we are ready to prove the converse.

\subsection*{Proof of Converse of Theorem \ref{thm:download}}
Starting from $k=2$ and applying Lemma \ref{lemma:recursive} repeatedly for $k = 3$ to $K$,
\begin{eqnarray}
&& \lefteqn{I(W_{2:K}; Q_{1:N}^{[1]}, A_{1:N}^{[1]} | W_1) }\nonumber\\
&\geq&  \frac{L}{N}(1-o(L)) + \frac{1}{N}I(W_{3:K};Q_{1:N}^{[2]}, A_{1:N}^{[2]} | W_1,W_2) \\
&\geq&  \frac{L}{N}(1-o(L)) + \frac{1}{N}\left[\frac{L}{N}(1-o(L)) + \frac{1}{N} I(W_{4:K};Q_{1:N}^{[3]}, A_{1:N}^{[3]} | W_{1:3}) \right] \\
&=&  {L}(1-o(L))(\frac{1}{N}+\frac{1}{N^2}) + \frac{1}{N^2} I(W_{4:K};Q_{1:N}^{[3]}, A_{1:N}^{[3]} | W_{1:3})\\
&\geq&  \cdots\\
&\geq&  {L}(1-o(L))(\frac{1}{N}+\cdots+\frac{1}{N^{K-2}}) + \frac{1}{N^{K-2}} I(W_{K};Q_{1:N}^{[K-1]}, A_{1:N}^{[K-1]} | W_{1:K-1})\\
&\geq&  {L}(1-o(L))(\frac{1}{N}+\cdots+\frac{1}{N^{K-1}})  \label{eq:recursive}
\end{eqnarray} 

\noindent Combining Lemma \ref{lemma:int} and (\ref{eq:recursive}), we have
\begin{eqnarray}
L(\frac{1}{R} - 1 + o(L)) \geq  {L}(1-o(L))(\frac{1}{N}+\cdots+\frac{1}{N^{K-1}})
\end{eqnarray}
Dividing both sides by $L$ and letting $L$ go to infinity gives us
\begin{eqnarray}
\frac{1}{R} - 1 &\geq&  \left(\frac{1}{N}+\cdots+\frac{1}{N^{K-1}}\right)\\
\Rightarrow R&\leq&\left(1+\frac{1}{N}+\cdots+\frac{1}{N^{K-1}}\right)^{-1}
\end{eqnarray}
thus, completing the proof.

\section{Discussion} \label{sec:disc}
In this section we share some interesting insights  beyond the capacity characterization.
\subsubsection*{\bf Upload Cost} To ensure privacy, we appealed to randomization arguments. To specify the randomly chosen query to the databases incurs an upload cost. For large messages the upload cost is negligible relative to the download cost, so it was ignored in this work. However, if the upload cost is a concern then it could be optimized as well.   Random permutations of message bits 
are sufficient for privacy, but it is easy to see that the upload cost can be reduced by reducing the number of possibilities to be considered. For example, consider the $K=2$ messages, $N = 2$ databases setting. We can group the bits, i.e., we can divide the 4 bits of each message into 2 groups, so that when we choose 2 bits, we only choose 2 bits from the same group. This reduces the choice to 1 out of 2 groups (rather than $2$ out of $4$ bits). Further, it may be possible to avoid random permutations among the chosen bits (group). For the same $K = 2$ messages and $N = 2$ databases example, we can fix the order within each group and the scheme becomes the following. We denote the messages bits as $W_1 = \{u_1, u_2, u_3, u_4\}, W_2 = \{v_1, v_2, v_3, v_4\}$.

\begin{eqnarray*}
\centering
\begin{array}{|c|c|c|c|c|}\hline
&\multicolumn{2}{c}{\mbox{Prob. $1/2$}} \vline&\multicolumn{2}{c}{\mbox{Prob. $1/2$}} \vline\\ \cline{2-5}
    & \mbox{Want $W_1$}  &\mbox{Want $W_2 $}& \mbox{Want $W_1$} &\mbox{Want $W_2$} \\
   \hline		
\mbox{Database $1$}&u_1, v_1, u_2 + v_2 &u_1, v_1, u_2 + v_2& u_3, v_3, u_4 + v_4&u_3, v_3, u_4 + v_4\\ \hline
\mbox{Database $2$}& u_4, v_2, u_3 + v_1 &u_2, v_4, u_1 + v_3&u_2, v_4, u_1 + v_3&u_4, v_2, u_3 + v_1\\ \hline
\end{array}
\end{eqnarray*}
Note that regardless of which message is desired, the user is equally likely to request either $u_1, v_1, u_2+v_2$ or $u_3,v_3,u_4+v_4$ from DB1, and either $u_2,v_4,u_1+v_3$ or $u_4,v_2,u_3+v_1$ from DB2, so the scheme is private. However, each query is now limited to only 2 possibilities, thereby significantly reducing the upload cost. Also note that instead of storing all $8$ bits that constitute the two messages, each database only needs to store $6$ bits in this case, corresponding to the two possible queries that it may face. Reducing the \emph{storage overhead} is an interesting question that has been explored by Fazeli, Vardy and Yaakobi in \cite{Fazeli_Vardy_Yaakobi}.

Another interesting question in this context is to determine the \emph{upload constrained capacity}. An information theoretic perspective is still useful. For example, since we are able to reduce the upload cost for $K=2, N=2$ to two possibilities, one might wonder if it is possible to reduce the upload cost of the $K=3,N=2$ setting to $3$ possibilities without loss of capacity. Let us label the three possible downloads from DB1 as $f_1, f_2, f_3$ and the three possible downloads from DB2 as $g_1, g_2, g_3$. We wish to find out if the original PIR capacity of $4/7$ is still achievable under these upload constraints. As we show next, the capacity is strictly reduced. With uploads limited to choosing one out of only 3 possibilities, the upload constrained capacity of the $K=3, N=2$ setting is $1/2$ instead of $4/7$. Eliminating trivial degenerate cases, in this case there is no loss of generality in assuming that we can recover $W_1$ from any one of these three possibilities: $(f_1, g_1), (f_2,g_2),(f_3,g_3)$; we can recover $W_2$ from any one of these three possibilities: $(f_1,g_2),(f_2,g_3),(f_3,g_1)$; and we can recover $W_3$ from any one of these three possibilities: $(f_1,g_3),(f_2,g_1),(f_3,g_2)$.  Then, for the optimal scheme we have
\begin{eqnarray}
H(W_1)&=&I(W_1;f_1,g_1)\\
&\leq&2H(A)-H(f_1,g_1|W_1)\\
\mbox{Similarly, }H(W_1)&\leq&2H(A)-H(f_2,g_2|W_1)\\
\mbox{Adding the two, } 2H(W_1)&\leq& 4H(A)-H(f_1,g_1,f_2,g_2|W_1)\\
&\leq&4H(A)-H(W_1,W_2,W_3|W_1)\label{eq:comp}\\
&\leq&4H(A)-H(W_2,W_3)\\
\Rightarrow C&=&H(W_1)/2H(A)\leq 1/2
\end{eqnarray}
Here, $2H(A)$ is the total download. (\ref{eq:comp}) follows because from $f_1,g_1,f_2,g_2$ we can recover all three messages. Thus, if the upload can only resolve one out of three possibilities for the query to each database, then the capacity of such a PIR scheme cannot be more than $1/2$, which is strictly smaller than the PIR capacity without upload constraints, $4/7$. In fact, the upload constrained capacity in this case is exactly $1/2$, as shown by the following achievable scheme which is interesting in its own right for how it fully exploits interference alignment. Suppose $W_1, W_2, W_3$ are symbols from a sufficiently large finite field (e.g., $\mathbb{F}_5$). Then the following construction works.
\begin{eqnarray}
f_1&=&W_1+2W_2+W_3\\
f_2&=&W_1+4W_2+3W_3\\
f_3&=&3W_1+4W_2+6W_3\\
\bigskip
g_1&=&W_1+4W_2+2W_3\\
g_2&=&3W_1+4W_2+3W_3\\
g_3&=&2W_1+4W_2+6W_3
\end{eqnarray}
It is easy to verify that $W_1$ can be recovered from any one of $(f_1, g_1), (f_2,g_2),(f_3,g_3)$; $W_2$ can be recovered from any one of $(f_1,g_2), (f_2,g_3),(f_3,g_1)$; and $W_3$ can be recovered from any one of $(f_1,g_3), (f_2,g_1), (f_3,g_2)$. The reason we can recover the desired message symbol from two equations, even though all three message symbols are involved in those two equations, is because of this special construction, which forces the undesired symbols to align into one dimension in every case. Thus, the upload constrained capacity for $K=3, N=2$ when the randomness is limited to choosing one out of $3$ possibilities, is $1/2$. Answering this question for arbitrary $K, N$ and arbitrary upload constraints is an interesting direction for future work.

\subsubsection*{Message Size} The information theoretic formulation of the PIR problem allows the sizes of messages to grow arbitrarily large. A natural question is this -- how large do we need each message to be for the optimal scheme. 
In our scheme, each message consists of $N^{K}$ bits.  However, even for our capacity achieving PIR scheme, the size of a message may be reduced. As an example, for the same $K = 2$ messages and $N = 2$ databases setting, the following PIR scheme works just as well (still achieves the same capacity) 
when each message is only made up of $2$ bits: $W_1=(u_1,u_2)$, $W_2=(v_1,v_2)$. 
\begin{eqnarray*}
\begin{array}{|c|c|c|c|c|}\hline
&\multicolumn{2}{c}{\mbox{Prob. 1/2}} \vline&\multicolumn{2}{c}{\mbox{Prob. 1/2}} \vline\\ \cline{2-5}
    & \mbox{Want $W_1 $} &\mbox{Want $W_2 $}& \mbox{Want $W_1$} &\mbox{Want $W_2$} \\
   \hline		
\mbox{Database 1}&u_1, v_2 &u_1, v_2& u_2, v_1&u_2, v_1\\ \hline
\mbox{Database 2}& u_2+v_2 &u_1+v_1& u_1 + v_1&u_2 + v_2\\ \hline
\end{array}
\end{eqnarray*}
Determining the smallest message size needed to achieve the PIR capacity, or the message size constrained PIR capacity, is another interesting direction for future work.

\subsection*{Similarities between PIR and Blind Interference Alignment}
The idea of blind interference alignment was introduced in \cite{Jafar_corr} to take advantage of the diversity of coherence intervals that  may arise in a wireless network. For instance, different channels may experience different coherence times and coherence bandwidths. A diversity of coherence patterns can also be artificially induced by the switching of reconfigurable antennas in pre-determined patterns. As one of the simplest examples of BIA, consider a $K$ user interference channel, where the desired channels have coherence time $1$, i.e., they change after every channel use, while the cross channels (which carry interference) have coherence time $2$, i.e., they remain unchanged over two channel uses. The transmitters are aware of the coherence times but otherwise have no knowledge of the channel coefficients. The BIA scheme operates over two consecutive channel uses. Over these two channel uses, each transmitter repeats its information symbol, and each receiver simply calculates the difference of its received signals. Since the transmitted symbols remain the same and the cross channels do not change, the difference of received signals from the two channel uses eliminates all interference terms. However, because the desired channels change, the desired information symbols survive the difference at each receiver. Thus, one desired information symbol is successfully sent for each message over 2 channel uses, free from interference, achieving $\frac{1}{2}$ DoF per message. Remarkably, this is essentially identical to the example of PIR included in the introduction, i.e., (\ref{eq:PIR1}), (\ref{eq:PIR2}). Applications of BIA extend well beyond this simple example \cite{Wang_Gou_Jafar, BIA_Dynamic, BIA_IC, BIA_Z}. For instance, in the  $X$ channel comprised of $M$ transmitters and $K$ receivers, using only the knowledge of suitable channel coherence patterns, BIA schemes achieve $\frac{MK}{M+K-1}$  DoF, which cannot be improved upon even with perfect channel knowledge  \cite{Jafar_corr, Wang_Gou_Jafar}. The connection to PIR also extends naturally as follows. 

The number of users in the BIA problem translates into the number of messages in the PIR problem. 
The received signals for user $\theta$  in BIA,  translate into the answering strings when message $W_\theta$ is the desired message in the PIR problem. The channel vectors associated with user $\theta$ in the BIA problem translate into the query vectors for desired message $W_\theta$ in the PIR problem. The privacy requirement of the PIR  scheme takes advantage of the observation that in BIA, over each channel use, the received signal at each receiver is statistically equivalent, because the transmitter does not know the channel values and the channel to each receiver has the same distribution. The most involved aspect of translating from BIA to PIR is that in BIA, the knowledge of the channel realizations across channel uses reveals the switching pattern, which in turn reveals the identity of the receiver. To remove this identifying feature of the BIA scheme, the channel uses are  divided into subgroups such that the knowledge of the switching pattern within each group reveals nothing about the identity of the receiver. Each sub-group of channel uses is then associated with a different database. Since the databases are not allowed to communicate with each other, and each sub-group of queries (channel uses) reveals nothing about the message (user), the resulting scheme guarantees privacy. {\color{black}Finally, the symmetric degrees of freedom (DoF) value per user in BIA is the ratio between the number of desired message symbols and the number of channel uses (received signal equations), and the rate $R$ in PIR is the ratio between the number of symbols of the desired message and the total number of equations in all answering strings. In this way, the DoF value achieved with BIA translates into the rate of the corresponding PIR protocol, i.e., $R = \mbox{DoF}$. We summarize these connections in the following table.}


\begin{table}[h]
\centering
\scalebox{1}{
\begin{tabular}{c | c}
   PIR & BIA \\
   \hline		
  Message & Receiver \\
  Queries &  Channel Coefficients \\
  Answers & Received Signals \\
  Rate & DoF
\end{tabular}
}
\end{table}

Recognizing this connection between PIR and BIA directly leads to capacity achieving PIR schemes for $K=2$ messages, and arbitrary number of databases $N$, as in \cite{Sun_Jafar_BIAPIR}, by translating from known optimal BIA schemes. However, for $K>2$, the PIR framework generalizes the BIA framework. This is because the coherence patterns that are assumed to exist in BIA are typically motivated by the distinct coherence times, coherence bandwidths, or antenna switching patterns that are feasible in wireless settings. However, since PIR is not bound by wireless phenomena, it allows for arbitrary coherence patterns, including many possibilities that would be considered infeasible in wireless settings. Even the simple scheme of BIA for the $K$ user interference channel presented earlier, was originally noted in BIA  \cite{Jafar_corr} merely as a matter of curiosity rather than having any physical significance. As such, while our initial insights into PIR came by viewing it as a special case of existing BIA schemes, the new capacity achieving PIR schemes introduced in this work go well beyond existing results in BIA, by allowing arbitrary coherence patterns.



\section{Conclusion}\label{sec:conc}
Information theorists commonly study the optimal coding rates of communication problems dealing with a few messages, each carrying an asymptotically large number of bits, while computer scientists often study the computational complexity of problems dealing with an asymptotically large number of messages, each carrying only a few bits (e.g., 1 bit per message). The occasional crossover of problems between the two fields opens up exciting opportunities for new insights. A prominent example is the index coding problem \cite{Birk_Kol, Birk_Kol_Trans}, originally posed by computer scientists and recently studied from an information theoretic perspective. The information theoretic capacity characterization for the index coding problem is now recognized as perhaps one of the most important open problems in network information theory, because of its fundamental connections to a broad range of questions that includes topological interference management, network coding, distributed storage, hat guessing, and non-Shannon information inequalities. Like index coding, the PIR problem also involves non-trivial interference alignment principles and is related to problems like blind interference alignment \cite{Jafar_corr} that have previously been studied in the context of wireless networks. In fact, it was the pursuit of these connections that brought us to the PIR problem \cite{Sun_Jafar_BIAPIR}. Further, PIR belongs to another rich class of problems studied in computer science, with deep connections to oblivious transfer \cite{SymPIR}, instance hiding \cite{Hide, Hide_one, Hide_multiple}, and distributed computation with untrusted servers \cite{Local_random}. Bringing this class of problems into the domain of information theoretic studies holds much promise for new insights and fundamental progress. The characterization of the information theoretic capacity of Private Information Retrieval is a step in this direction. 

\bibliographystyle{IEEEtran}
\bibliography{Thesis}
\end{document}